\documentclass[journal]{IEEEtran}
\usepackage{amsmath,amsfonts}
\usepackage{algorithmic}
\usepackage{algorithm}
\usepackage{array}
\usepackage[caption=false,font=normalsize,labelfont=rm,textfont=rm]{subfig}
\usepackage{textcomp}
\usepackage{stfloats}
\usepackage{url}
\usepackage{verbatim}
\usepackage{graphicx}
\usepackage{cite}
\usepackage{balance}
\usepackage{subcaption}
\usepackage{bm}
\usepackage{orcidlink}
\usepackage{makecell}
\usepackage{hyperref}
\usepackage{xr-hyper}
\newtheorem{proposition}{Proposition}

\begin{document}

\title{Multi-Level Damage-Aware Graph Learning for Resilient UAV Swarm Networks}

\author{Huan Lin~\orcidlink{0009-0002-7217-6139}, Chenguang Zhu~\orcidlink{0000-0002-6790-7091}, Lianghui Ding~\orcidlink{0000-0002-3231-3613 },~\IEEEmembership{Member,~IEEE}, Lin Wang~\orcidlink{0000-0002-8374-8473}, Feng Yang~\orcidlink{0000-0002-6350-5765}
\thanks{This paper was supported in part by Shanghai Key Laboratory Funding under Grant STCSM15DZ2270400, and in part by the Program for Professor of Special Appointment (Eastern Scholar) at Shanghai Institutions of Higher Learning. (\textit{Corresponding author: Lianghui Ding.})}
\thanks{Huan Lin, Chenguang Zhu, and Lianghui Ding are with the Institute of Image Communication and Network Engineering, School of Integrated Circuits, Shanghai Jiao Tong University, Shanghai 200240, China. Email: lhzt715@sjtu.edu.cn; n1cex9@sjtu.edu.cn; lhding@sjtu.edu.cn.}
\thanks{Lin Wang is with the State Key Laboratory of Submarine Geoscience, School of Automation and Intelligent Sensing, Shanghai Jiao Tong University, Shanghai 200240, China. Email: wanglin@sjtu.edu.cn.}
\thanks{Feng Yang is with the Institute of Wireless Communication Technologies, School of Integrated Circuits, Shanghai Jiao Tong University, Shanghai 200240, China. Email: yangfeng@sjtu.edu.cn.}

}



\maketitle

\begin{abstract}
Unmanned aerial vehicle (UAV) swarm networks leverage resilient algorithms to restore connectivity from communication network split issues. However, existing graph learning-based approaches face over-aggregation and non-convergence problems caused by uneven and sparse topology under massive damage. 
In this paper, we propose a novel Multi-Level Damage-Aware (MLDA) Graph Learning algorithm to generate recovery solutions, explicitly utilizing information about destroyed nodes to guide the recovery process. The algorithm first employs a Multi-Branch Damage Attention (MBDA) module as a pre-processing step, focusing attention on the critical relationships between remaining nodes and destroyed nodes in the global topology. By expanding multi-hop neighbor receptive fields of nodes to those damaged areas, it effectively mitigating the initial sparsity and unevenness before graph learning commences.
Second, a Dilated Graph Convolution Network (DGCN) is designed to perform convolution on the MBDA-processed bipartite graphs between remaining and destroyed nodes. The DGCN utilizes a specialized bipartite graph convolution operation to aggregate features and incorporates a residual-connected architecture to extend depth, directly generating the target locations for recovery. 
We theoretically proved the convergence of the proposed algorithm and the computational complexity is acceptable. Simulation results show that the proposed algorithm can guarantee the connectivity restoration with excellent scalability, while significantly expediting the recovery time and improving the topology uniformity after recovery.
\end{abstract}


\section{Introduction}
\IEEEPARstart{T}{echnology} of unmanned aerial vehicle (UAV) has developed rapidly in recent years and has been widely applied with the advantages of low cost and high flexibility. Meanwhile, to overcome the limited capability of single UAVs, the UAV swarm composed of hundreds of UAVs has been developed to conduct various complex tasks in open and dangerous environments, such as rescue \cite{rescue}, reconnaissance \cite{reconnaissance}, target attack \cite{attack}, etc. 
The UAV swarm usually forms an ad-hoc network for collaboration, where task and control information is transmitted between UAVs through wireless communication links. However, the UAV swarm network (USNET) is susceptible as the UAVs face hostile strikes and external destruction. 
In the worst case, the damage of multiple UAVs can divide the USNET into isolated sub-nets and thus cause the communication network split (CNS) issue.
Several pro-active resilient mechanisms were proposed to reduce the probability of CNS issues via topology enhancement \cite{active1, active2, active3}, yet these strategies can not maintain the connectivity under severe damage scenarios. Therefore, the USNET requires reactive resilient algorithms to restore the connectivity in minimum recovery time.

Resilient algorithms were first studied in wireless sensor networks (WSNs) under static scenarios \cite{wsn1, wsn2, wsn3, wsn4}. Subsequently, with the advent of mobile WSNs, mobility-based algorithms were developed to restore communication connectivity through node movements \cite{mv_cut1, mv_cut2, mv_cut3, mv_ms1, mv_wsn1, mv_wsn2, mv_wsn3, mv_opt1, mv_opt2, mv_opt3}, and were later extended for USNETs \cite{hero, sidr, mgc, demd}. These approaches can be categorized into critical node replacement, central aggregation, and optimal planning algorithms. The first category moves the backup neighboring nodes to replace the position of destroyed critical nodes \cite{mv_cut1, mv_cut2, mv_cut3, mv_ms1}, yet they suffer from severe cascaded motion problems under multiple destroyed UAVs. The second category aggregates the remaining nodes toward a central meeting point \cite{mv_wsn1, mv_wsn2, mv_wsn3, hero, sidr}, whereas these algorithms often result in prolonged recovery time, as the time is determined by the node farthest from the meeting point.
The third category focuses on minimizing the recovery time by planning optimal recovery trajectories for every remaining node \cite{mv_opt1, mv_opt2, mv_opt3}. Through the optimization of individual recovery paths, the algorithm ensures that each node moves along the most efficient route, reducing overall recovery time.

Among existing optimal planning algorithms, graph-learning-based approaches \cite{gat, mgc, demd} have achieved state-of-the-art performance by considering the interactions between UAV nodes and leveraging the powerful information aggregation capabilities of Graph Convolution Networks (GCN).
However, there are three major challenges in using graph-learning algorithms to address CNS issues. 
The first is spatial over-aggregation that arises from the uneven topology of the remaining network, where certain nodes become highly clustered. This can lead to some nodes focusing excessively on a few high-degree neighbors while neglecting others, resulting in inefficient recovery trajectories that aggregate around high-degree nodes.
The second challenge is that a sparse network topology under extensive damage can impede the convergence of graph-based algorithms, as nodes lack sufficient local connectivity to effectively propagate information. Consequently, these algorithms may fail to restore the connectivity.
The third challenge lies on the scalability. Since GCN is a transductive learning approach \cite{gcn}, the changes of damage cases and swarm scales require additional online iterations or numerous pre-trained models, harming the deployment feasibility of graph learning-based algorithms.

In this paper, we propose a Multi-Level Damage-Aware Graph Learning (ML-DAGL) algorithm that leverages location of destroyed UAVs to guide the recovery process, which generates recovery trajectories through two steps.
First, a Multi-Branch Damage Attention (MBDA) module forms a series of multi-hop Damage-Attentive Graphs (mDAGs) that can overcome all three problems outlined above. By assigning the attention of surviving nodes to their destroyed neighbors, mDAGs form more even topologies and help mitigate spatial over-aggregation. The multi-hop dilation of mDAGs also addresses sparse topology issues by improving nodes' receptive fields in severe damage regions.
Moreover, the unified structure of mDAGs is insensitive to the scale of damage as they always considers all nodes, thus they have good scalability under damage-varying cases.
Second, due to the special bipartite topology of mDAGs, we design a Dilated Graph Convolution Network (DGCN) to compute recovery trajectories based on mDAGs. The Bipartite Graph Convolution Operation (B-GCO) in DGCN efficiently aggregates features among remaining and destroyed nodes, and the residual connection mechanism preserves the gradient for deep structure.
Theoretical analysis showed that the proposed algorithm has stable convergence and acceptable computational complexity.
Simulation results indicate that the proposed algorithm guarantees connectivity restoration across varying damage scenarios, significantly reduces average recovery time, and enhances the recovered topology with a more uniform degree distribution.
Our contributions in this paper are summarized as follows.
\begin{enumerate}
    \item We propose a novel ML-DAGL algorithm for UAV swarm connectivity restoration, explicitly incorporating all UAV nodes, both surviving and destroyed, into a fixed-dimension representation as mDAG. This fundamental design decouples model input from specific damage patterns, enabling a single pre-trained model to handle any damage scale within a fixed swarm scale, directly solving the scalability challenge of GCNs.
    \item We introduce the MBDA module to pre-process the global topology, with a novel damage attention mechanism that focus on critical surviving-destroyed node links. Combined with parallel multi-hop dilated processing, it actively reshapes sparse and uneven topologies into dense, uniform mDAGs before learning, inherently mitigating over-aggregation and information sparsity.
    \item We design DGCN to operate on the mDAGs. It uses a bipartite graph convolution for feature aggregation from destroyed nodes to surviving nodes and a residual architecture for depth. Its theoretical convergence proof ensure stable training and reliable trajectory generation even under severe fragmentation, resolving the non-convergence challenge of prior GCNs.
\end{enumerate}

The remainder of this paper is organized as follows. Section \uppercase\expandafter{\romannumeral2} analyzes the related work. Section \uppercase\expandafter{\romannumeral3} forms the system model of the USNET graph and CNS issue. Section \uppercase\expandafter{\romannumeral4} describes the
specific details of the proposed algorithm, and experiment results are given in Section \uppercase\expandafter{\romannumeral5}. Finally, Section \uppercase\expandafter{\romannumeral6} presents the conclusion.

\section{Related Work}
Resilient recovery algorithms for USNET have become a critical area of research, particularly in scenarios where connectivity is disrupted due to massive node destruction in hostile environments \cite{survey}. Various strategies have been developed to address the CNS challenges, focusing on reconnecting divided sub-nets within minimum recovery time. 
Considering the motivations for moving the surviving nodes, we can divide mobility-based approaches into critical node replacement, central aggregation, and optimal planning algorithms.

The critical-node replacement algorithms aim to rebuild connectivity by replacing the locations of destroyed critical nodes with the backup neighboring nodes. For example, algorithms in \cite{mv_cut1, mv_cut2} identified the global cut-vertex nodes as the critical nodes and addressed the CNS issues in small networks. Zhang \textit{et al.}\cite{mv_cut3} extended the recovery algorithm to large-scale WSNs by employing a finite state machine, effectively reducing the complexity of identifying cut-vertex nodes. Younis \textit{et al.}\cite{mv_ms1} considered a clustered USNET and designated the master nodes as the critical nodes, eliminating the need for additional identification processes. However, the cascaded motion problem under massive damage will increase the moving distances of node replacement and reduce the network's coverage ability \cite{mv_ms2}. Therefore, the node-replacement algorithms show poor performance in addressing CNS issues under massive damage scenarios.

The central aggregation algorithms guide the remaining nodes towards a pre-defined meeting point, which can ensure the restoration of connectivity. These algorithms have been widely applied in mobile WSNs \cite{mv_wsn1, mv_wsn2, mv_wsn3}. In the field of USNETs, Mi \textit{et al.}\cite{hero} proposed to define multiple meeting points, which reduces the average moving distance of edge nodes. Chen \textit{et al.}\cite{sidr} developed a swarm intelligence-based damage-resilient mechanism that calculates the recovery path for each sub-network as an entire part. This approach allows the network to restore connectivity without losing its original links and achieves path tracking with low communication overhead. 
However, because all UAVs aggregate towards a single meeting point, the network's recovery time is determined by the node farthest from the meeting point. Consequently, central aggregation methods typically result in longer recovery time costs compared to the other two categories of algorithms.

The optimal-planning algorithms formulate the recovery issues as an optimization problem to minimize the recovery time via planning optimal recovery paths for the remaining nodes \cite{mv_opt1, mv_opt2, mv_opt3}. With the development of UAV intelligence, more advanced approaches integrate deep learning and graph-based algorithms to evolve the network topologies. Mou \textit{et al.}\cite{mgc} applied the graph convolution network (GCN) to solve the CNS challenge for the first time, where the information aggregation capability of graph learning enables the algorithm to achieve excellent performance on shorter recovery time. Our previous work \cite{demd} developed a damage-embedding module that maps the location information of destroyed nodes into the features of their remaining neighbors. This mechanism strengthens the GCN's capability of modeling and reasoning, improving the performance by inputting more adequate information.
However, the uneven topology of the remaining network led to a serious over-aggregation problem as the nodes were attracted to their high-degree neighbors, and the sparse topology after severe damage ultimately degraded the performance of graph learning.

This paper aims to explore a more efficient graph-learning algorithm to eliminate the problems mentioned above. By re-attracting the attention of the remaining nodes to damaged nodes with topology processing, we consider forming a graph with only links between remaining and damaged nodes to avoid the uneven topology and utilizing multi-hop dilation to enhance the sparse topology, thereby achieving fast connectivity restoration and more robust recovered networks. 

\section{System model}
This section first constructs a graph topology to represent the UAV swarm network (USNET) and then proposes the massive damage model. The problem investigated in this paper is formulated in the third subsection.

\subsection{UAV Communication Model}
Assuming that the USNET is composed of $N$ massive independent and identical UAVs with a pre-defined connected structure. Generally, UAVs form a network at equal altitudes for most missions, hence we mainly consider two-dimensional locations. The location vector $\bm{p}_i(t)={[x_i(t),y_i(t)]}^\top \in \mathbb{R}^2$ denotes the coordinates of the $i$-th UAV at time $t$, where $x_i(t)$ and $y_i(t)$ denote the coordinate component of $X$ and $Y$ axis, respectively.

Since we focus on the CNS issue in USNETs, the transmission delay of the air-to-air communication channel is not considered in this paper. Without loss of generality, we consider a predominantly line-of-sight communication link between UAVs in the air. According to Friis transmission formula, the received signal power of $u_j$ from $u_i$ is
\begin{align}
    P_{ij}=P_0G_{tx}G_{rx}L(d_{ij})|h_0|^2,
\end{align}
where $P_0$ is the transmitting power, $G_{tx},G_{rx}$ are the constant gains of the transmitting and receiving antennas, respectively. $L(d_{ij})=\left(\lambda_c/{4\pi d_{ij}}\right)^2$ is the large-scale fading, where $\lambda_c$ is the wavelength, $d_{ij}=\|\bm{p}_i-\bm{p}_j\|$ is the Eulerian distance between the UAV $u_i$ and $u_j$. The $|h_0|^2$ is the small-scale fading that follows the Gamma distribution. Denoting the receiver sensitivity $\tau$, condition of UAVs $u_i$ and $u_j$ that can form a communication link follows
\begin{align}
    P_0G_{tx}G_{rx}L(d_{ij})|h_0|^2\leq\tau.
    \label{dtr}
\end{align}

We can then determine the maximum transmission range $d_{tr}$ of UAV by solving (\ref{dtr}) as an equation. Therefore, the communication link between $u_i$ and $u_j$ exists if and only if $d_{ij}(t)\leq d_{tr}$.

\subsection{Graph Topology of USNET}
Let graph $\mathcal{G}(t)=\{\mathcal{U}(t),\mathcal{E}(t)\}$ denote the topology of the USNET, where $\mathcal{U}(t)=\{u_i|i=1,2,...,N\}$ represents the node set and $\mathcal{E}(t)=\{e_{ij}|u_i,u_j\in \mathcal{U}(t)\}$ represents the edge set of communication links. The edge $e_{ij}$ exists if and only if node $u_i$ and $u_j$ form a communication link. Generally, the USNET establishes an ad hoc network to enable message exchange across the multi-hop links. Denoting $H_{ij}(t)$ as the number of hops between nodes $u_i$ and $u_j$ at time $t$, node $u_j$ is called a $k$-hop neighbor of $u_i$ at time $t$ if $H_{ij}(t)=k$.

With the help of satellite devices or ground stations, UAV nodes can easily obtain global information for local computing. In this paper, we assume that each UAV periodically reports its position and status information through satellite communication devices. Therefore, the satellite can record the node status as on-line and off-line, and summarize the global position  matrix $\bm{X}(t)$ consisting of the location vectors of each UAV at time $t$, i.e., 
\begin{align}
    \bm{X}(t)=[\bm{p}_1(t),\bm{p}_2(t),...,\bm{p}_N(t)]^\top \in \mathbb{R}^{N\times 2}.
\end{align}
The global information of node status and position matrix are then broadcasted to each UAV nodes periodically for local computation.

Notice that the damage can occur at any time during and destroy a certain number of UAVs, there can be more than $2^N$ different damage cases under varying numbers of destroyed UAVs. Since not every damage cases will split the network, we only study the damage scenarios in which can result in the communication network split (CNS) issues. The damage model in this paper is simplified as $N_D$ UAVs were destroyed at time $t_0$, and $N_R=N-N_D$ UAVs remained with CNS issues. Based on the status vector, we can label the damage index $\{d_1,d_2,...,d_{N_D}\}$ and the remain index $\{r_1,r_2,...,r_{N_R}\}$. To model the topology after the damage, we constructed a remaining graph $\mathcal{G}_R(t)=\{\mathcal{U}_R(t),\mathcal{E}_R(t)\}$ at time $t\geq t_0$, where $\mathcal{U}_R(t)=\{u_{r_i}|r_i=r_1,r_2,...,r_{N_R}\}$ is the remaining-node set, and $\mathcal{E}_R(t)=\{e_{r_ir_j}|e_{r_ir_j}\in\mathcal{E}(t), u_{r_i},u_{r_j}\in \mathcal{U}_R(t)\}$ is the edge set for remaining UAVs.

\subsection{Network Connectivity}
Due to the CNS issue, the remaining graph $\mathcal{G}_R(t)$ is split into several disconnected sub-nets.
Let $N_s(t)$ denote the number of sub-nets in $\mathcal{G}_R(t)$ at time $t$, apparently, $\mathcal{G}_R(t)$ is connected at time $t$ if and only if $N_s(t)=1$. 

A widely-used method to calculate $N_s(t)$ of $G_R(t)$ with the Laplace matrix is introduced in \cite{laplacian}. We first define the adjacent matrix of $\mathcal{G}(t)$ at time $t$ as $\bm{A}(t)=(a_{ij})\in \mathbb{S}^N$, where $a_{ij}$ is a 0-1 variable to denote if $e_{ij}$ exists or not, and $\mathbb{S}^N$ represents the set of $N\times N$ dimensional symmetric matrices. Similarly, the adjacent matrix of the $\mathcal{G}_R(t)$ at time $t$ is defined as $\bm{A}_R(t)=(a_{r_ir_j})\in \mathbb{S}^{N_R}$, and $a_{r_ir_j}=1$ if and only if $e_{r_ir_j}\in \mathcal{E}_R(t)$. With the adjacent matrix $\bm{A}_R(t)$, we can calculate degree matrix of $\mathcal{G}_R(t)$ as
\begin{equation}
\bm{D}_R(t)={\rm diag}(d_{r_1}(t),d_{r_2}(t),...,d_{r_{N_R}}(t)),
\label{degree}
\end{equation}
where $d_{r_i}(t)=\sum_{j=r_1}^{r_{N_R}}a_{r_ir_j}(t)$ is the degree of each remaining node, and ${\rm diag}(\cdot)$ represents a diagonal matrix. 

The Laplace matrix of $\mathcal{G}_R(t)$ is then calculated as the difference between $\bm{D}_R(t)$ and $\bm{A}_R(t)$, i.e., 
\begin{equation}
    \bm{L}_R(t)=\bm{D}_R(t)-\bm{A}_R(t).
    \label{laplace}
\end{equation}
Note that $\bm{L}_R(t)$ is a positive semi-definite matrix, with the eigenvalue decomposition applied on $\bm{L}_R(t)$, we have
\begin{equation}
    \bm{L}_R(t)=\bm{U}_R(t)\bm{\Lambda}_R(t)\bm{U}_R^\top(t),
\end{equation}
where $\bm{U}_R(t)=[\bm{u}_{r_1}(t),\bm{u}_{r_2}(t),...,\bm{u}_{r_{N_R}}(t)]$ as the unitary matrix composed of orthogonal eigenvectors, and $\bm{\Lambda}_R(t)={\rm diag}(\lambda_{r_1}(t),\lambda_{r_2}(t),...,\lambda_{r_{N_R}}(t))$ denotes the matrix with non-negative eigenvalues. The number of sub-nets is equal to the number of zero eigenvalues in $\bm{\Lambda}_R(t)$, i.e., 
\begin{align}
    N_s(t)={\rm Count}(\lambda=0|\bm{\Lambda}_R(t)).
\end{align}

\subsection{Problem Formulation}
The goal of resilient algorithms is to restore the connectivity of $\mathcal{G}_R(t)$, i.e., decreasing $N_S(t)$ to 1. Compared with computing instant velocities for each UAVs during the recovery process, a simplified way is to generate a target position matrix $\hat{\bm{X}}_R=[\hat{\bm{p}}_{r_1},\hat{\bm{p}}_{r_2},...,\hat{\bm{p}}_{r_{N_R}}]^\top$, which forms a target connected graph $\mathcal{\hat{G}}_R$ with $\hat{N}_S=1$. Therefore, the recovery process is presented as each node moves towards its target location straightly with a maximum speed $v_{max}$ then hovers in place.

We now define the recovery time cost $T_{rc}$ to measure the efficiency of algorithms. We neglect inertial effects as it is often done in the control literature on swarm resilience \cite{sidr,mgc}. The flying time of the remaining node $u_{r_i}$ is given by
\begin{align}
    T_{r_i}=\frac{1}{v_{max}}\|\bm{\hat{p}}_{r_i}-\bm{p}_{r_i}(t_0)\|. 
\end{align}
Obviously, the total recovery time $T_{rc}$ depends on the maximum flying time of the final remaining UAV reaching its target position, i.e., $T_{rc}={\rm max}_{u_{r_i}\in \mathcal{U}_R}T_{r_i}$. Take $T_{rc}$ as the optimization target along with the connectivity constraint, the problem of CNS recovery is formulated as 
\begin{align}
    ({\rm P}1):\quad \mathop{\rm min}\limits_{\bm{\hat{X}}_R}\quad & T_{rc}=\mathop{\rm max}\limits_{u_{r_i}\in\mathcal{U}_R}\frac{\|\bm{\hat{p}}_{r_i}-\bm{p}_{r_i}(t_0)\|}{v_{max}}
    \label{p2}\\
    \quad {\rm s.t.}\quad & \hat{N}_S = 1,
    \label{c2}
\end{align}


\begin{figure*}[!t]
\centerline{\includegraphics[width=\linewidth]{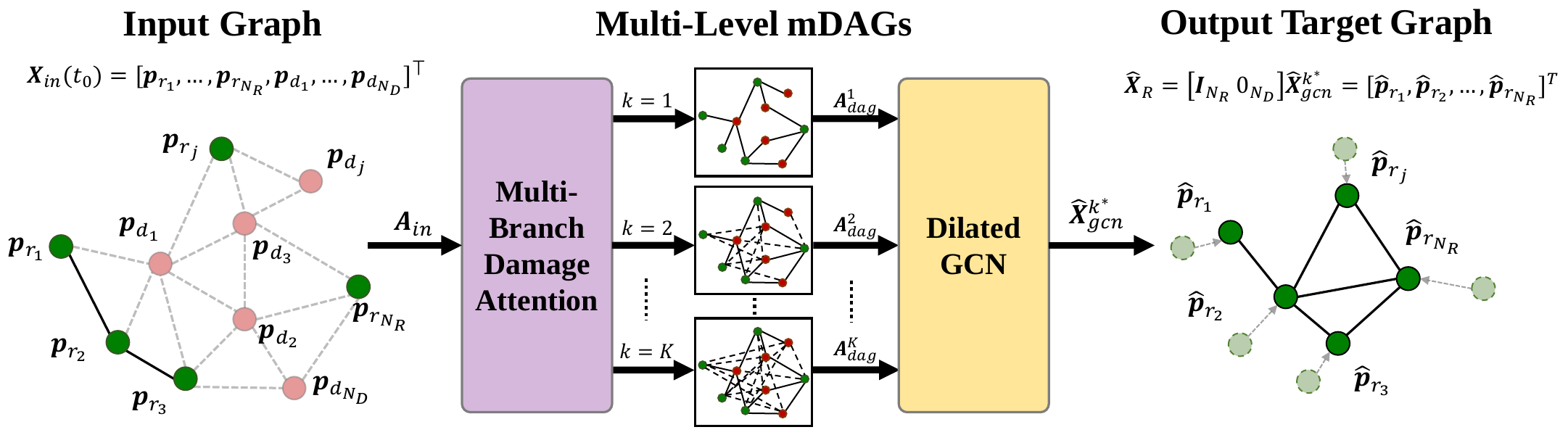}}
\captionsetup{justification=raggedright, singlelinecheck=false}
\caption{Overall framework of the proposed ML-DAGL algorithm.}
\label{framework}
\end{figure*}

\section{Approach}
This section first introduces the overall framework of the proposed ML-DAGL algorithm. Then, the details of the proposed MBDA module and DGCN are presented in the second and third subsections, respectively. Finally, the proposed algorithm is theoretically analyzed in the last subsection.


\subsection{Overall Framework of ML-DAGL}
As shown in Fig.\ref{framework}, our proposed ML-DAGL algorithm has an end-to-end architecture that can provide a target position matrix for the remaining graph to solve (P1). The ML-DAGL framework consists of two key components: a Multi-Branch Damage Attention (MBDA) module to form the multi-hop Damage Attentive Graphs (mDAGs), and a Dilated Graph Convolution Network (DGCN) to generate solutions based on mDAGs.

The proposed algorithm first constructs an input graph $\mathcal{G}_{in}(t_0)=\{\mathcal{U}_{in},\mathcal{E}_{in}\}$ to resort the order of UAVs based on node status, i.e., $\mathcal{U}_{in}=\{u_{r_1},...,u_{r_{N_R}},u_{d_1},...,u_{d_{N_D}}\}$. Then, we can determine the input feature matrix as the position matrix of $\mathcal{G}_{in}$, that is
\begin{align}
    \bm{X}_{in}(t_0)=[\bm{p}_{r_1}(t_0),...,\bm{p}_{r_{N_R}}(t_0),\bm{p}_{d_1}(t_0),...,\bm{p}_{d_{N_D}}(t_0)]^\top.
\end{align}
The adjacent matrix of $\mathcal{G}_{in}$ is calculated as $\bm{A}_{in}=(a_{in,ij})$ based on $\mathcal{E}_{in}$, where $a_{in,ij}=1$ if and only if link $e_{in,ij}$ exists.

The MBDA module transforms the input adjacent matrix $\bm{A}_{in}$ into several mDAGs via parallel branches. The $k$-th branch outputs an mDAG with its adjacent matrix $\bm{A}_{dag}^k$, which establishes links between remaining and damaged nodes within the range of $k$-hop neighborhood. With the number of branches $K$, the module finally forms $K$ different mDAGs $\{\bm{A}_{dag}^1,\bm{A}_{dag}^2,...,\bm{A}_{dag}^K\}$.

Then, DGCN applies convolution on the series of mDAGs and outputs a target position matrix $\bm{\hat{X}}_{gcn}^{k^*}$. Consequently, we can determine the optimal target position matrix for (P1) as
\begin{align}
    \bm{\hat{X}}_R=[\bm{I}_{N_R}\enspace\bm{0}_{N_D}]\bm{\hat{X}}_{gcn}^{k^*}=[\bm{\hat{p}}_{r_1}^{k^*},\bm{\hat{p}}_{r_2}^{k^*},...,\bm{\hat{p}}_{r_{N_R}}^{k^*}]^\top,
    \label{sl}
\end{align}
where $\bm{I}_{N_R}$ is the $N_R$-by-$N_R$ identity matrix, $\bm{0}_{N_D}$ is the $N_D$-by-$N_D$ zero matrix.

Since the proposed ML-DAGL is an optimization-based transductive algorithm, it requires additional online iterations to generate an optimal solution for a specific input graph $\mathcal{G}_{in}$. To reduce online iterations, pre-trained models are deployed to initialize the parameters of DGCN and skip the warm-up stage in training. Note that all mDAGs share an unified shape related to the number of UAVs $N$, a pre-trained model for a certain swarm scale can be adapted to distinct damage cases under varying $N_D$. In summary, the ML-DAGL requires limited online iterations and storage requirements, hence it is feasible for practical deployment on UAVs.

\subsection{Multi-Branch Damage Attention}
\begin{figure*}[!t]
\centerline{\includegraphics[width=.99\linewidth]{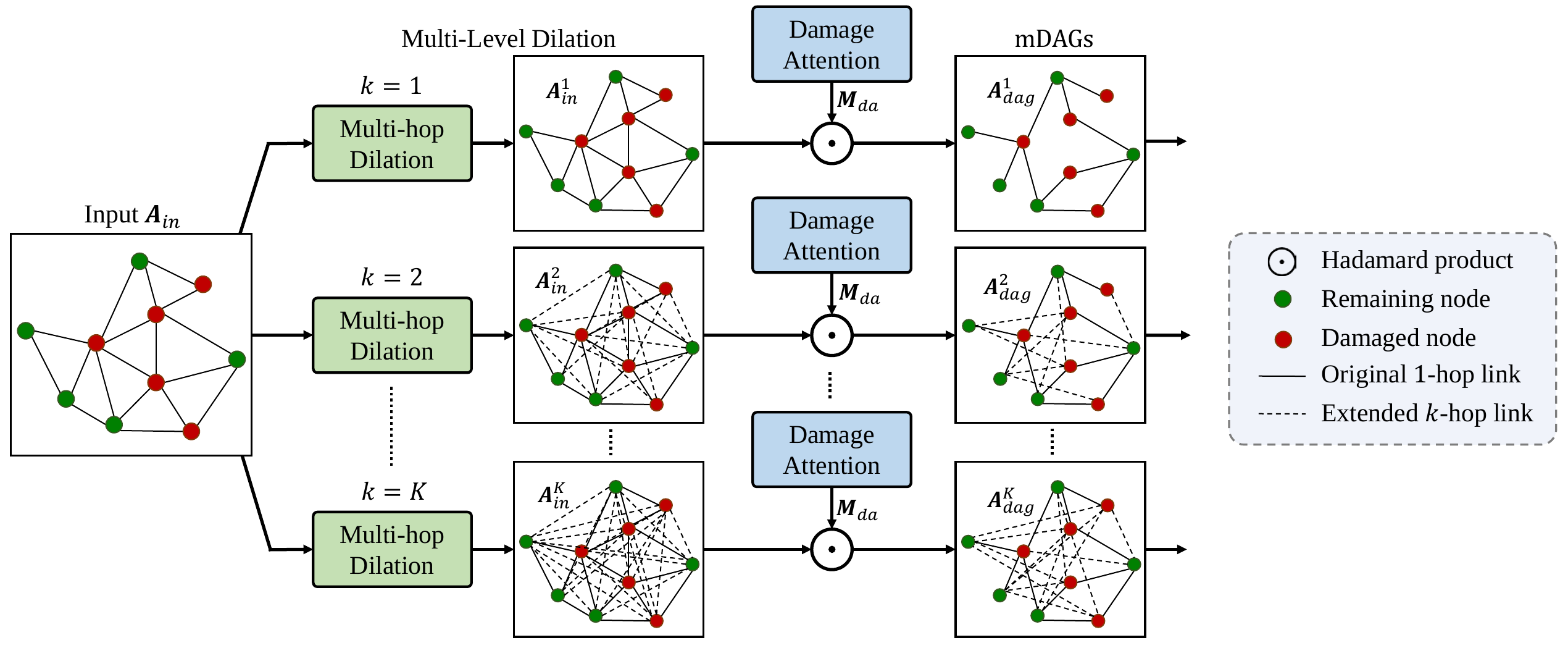}}
\captionsetup{justification=raggedright, singlelinecheck=false}
\caption{Structure of the proposed MBDA module.}
\label{mbda}
\end{figure*}

The destroyed neighbors of the remaining nodes can effectively depict the damage model within local areas and guide the remaining nodes in filling gaps on the damaged regions. To achieve this, we propose the MBDA module that extends the receptive fields via multi-hop dilation and applies damage attention to form mDAGs. The module structure is illustrated in Fig.\ref{mbda}.

\subsubsection{Multi-hop dilation branches}
To enhance the sparse topology of the input graph $\mathcal{G}_{in}$, we add edges between nodes and their multi-hop neighbors for receptive fields dilation. However, a static dilation size has limited performance when facing dynamic damage scales, since a large dilation size benefits the massive damage scenarios, while small-scale damage only requires a small dilation.
In our approach, we develop $K$ parallel dilation branches to provide mDAG with multi-level damage-aware regions. 

As shown in Fig.\ref{mbda}, the $k$-th branch applies $k$-hop dilation on  $\bm{A}_{in}$ and outputs the dilated adjacent matrix $\bm{A}_{in}^k=(a_{in,ij}^k)\in \mathbb{S}^N$. The element $a_{in,ij}^k$ is calculated by
\begin{equation}
    \begin{aligned}
        a_{in,ij}^k=
        \begin{cases}
            1,\quad  &H_{in,ij}\leq k;\\
            0,&otherwise,
        \end{cases}
    \end{aligned}
\end{equation}
where $H_{in,ij}$ denotes the number of hops between nodes $u_{in,i}$ and $u_{in,j}$ in $\mathcal{G}_{in}$. In other words, with the dilation size as $k$, $\bm{A}_{in}^k$ will form links between nodes whose number of hops is no larger than $k$. After the multi-hop dilation branches, we can get a series of dilated adjacent matrices $\{\bm{A}_{in}^1,\bm{A}_{in}^2,...,\bm{A}_{in}^K\}$. 

\subsubsection{Damage attentive graph}
To cope with the over-aggregation problem caused by the uneven topology after damage, we propose the damage attention that focuses on the relations between the remaining nodes and their destroyed neighbors, i.e., only links between the remaining nodes $u_{r_i}$ and destroyed nodes $u_{d_j}$ will be kept in the adjacent matrix. 

In the proposed MBDA shown in Fig.\ref{mbda}, we design a damage attention matrix $\bm{M}_{da}$ to transform the adjacent matrix, which is defined as
\begin{align}
    \bm{M}_{da} = 
    \begin{bmatrix}
    \bm{0}_{N_R} & \bm{J}_{N_R,N_D} \\
    \bm{J}_{N_R,N_D}^\top & \bm{0}_{N_D}
    \end{bmatrix},
\end{align}
where $\bm{J}_{N_R,N_D}$ is a $N_R$-by-$N_D$ ones matrix that all elements equal to 1.

The $k$-th branch then calculates the adjacent matrix of the mDAG as
\begin{align}
    \bm{A}_{dag}^k = \bm{A}_{in}^k\odot \bm{M}_{da} =
    \begin{bmatrix}
    \bm{0}_{N_R} & \bm{A}_{in,r_id_j}^k \\
    \bm{A}_{in,r_id_j}^{k\top} & \bm{0}_{N_D}
    \end{bmatrix},
    \label{Hadamard}
\end{align}
where $\bm{A}_{in,r_id_j}^k$ denotes the $k$-hop dilated links between $u_{r_i}$ and $u_{d_j}$, and $\odot$ denote the Hadamard product operator of matrices. Since nodes in $\bm{A}_{in,r_id_j}^k$ can be divided into the remaining-node and destroyed-node sets and no links exist between nodes within the same set, the mDAG is so-called a bipartite graph. This property splits the relationships between nodes distributed in uneven sub-nets, ensuring that the remaining nodes in the graph do not interact. After the damage attention processing, we have the mDAG sequence with $\{\bm{A}_{dag}^1,\bm{A}_{dag}^2,...,\bm{A}_{dag}^K\}$. 

The parallel structure of MBDA module allows the algorithm to always determine an optimal $k^*$ for each damage case, adapting to varying scales of swarm $N$ and damage $N_D$. While available computational resources constrain the choice of the number of branches $K$, later analysis will demonstrate that the parallel structure significantly reduces both pre-training overhead and storage requirements.

\subsection{Dilated Graph Convolution Network}
\begin{figure}[!t]
\centerline{\includegraphics[width=.98\linewidth]{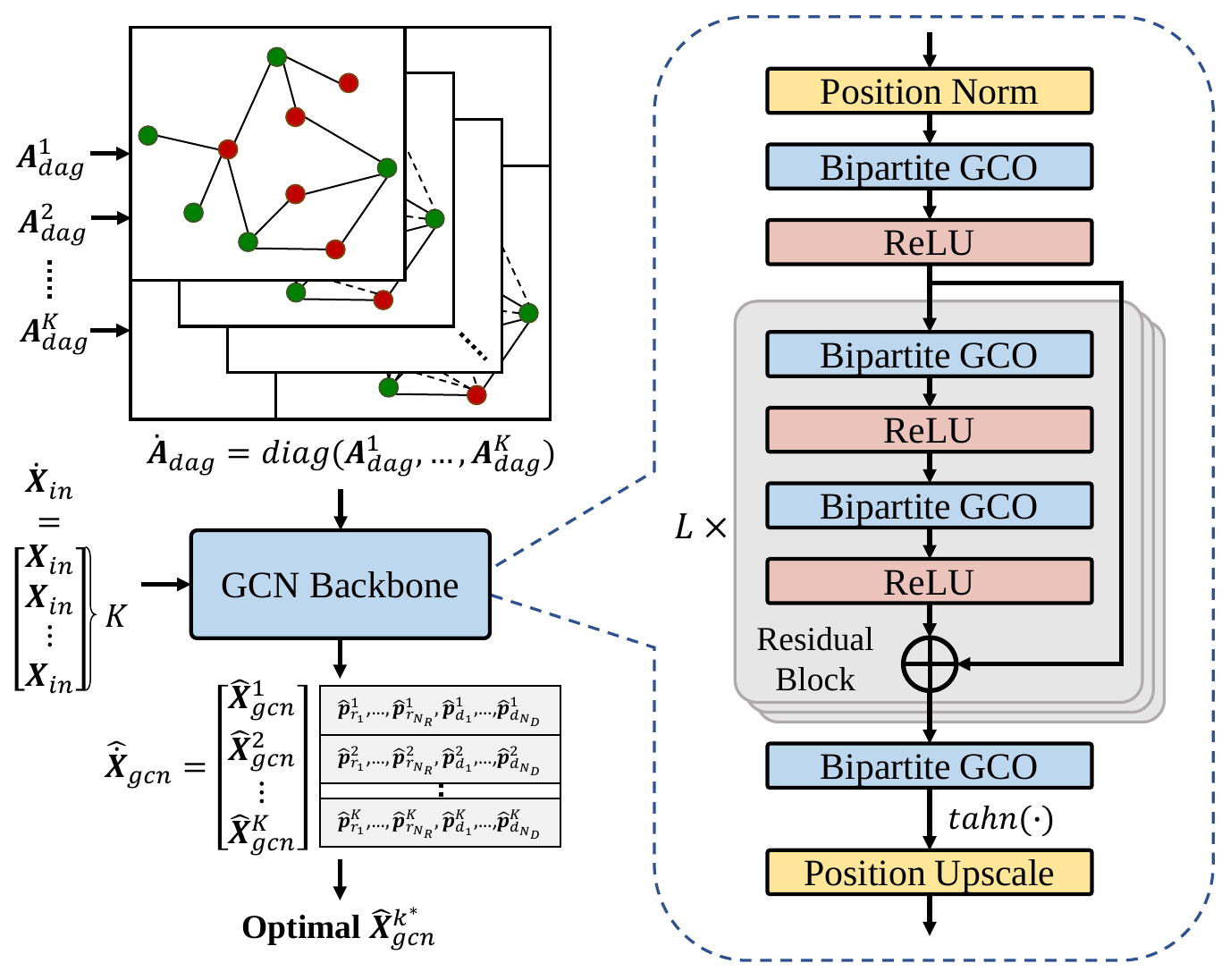}}
\captionsetup{justification=raggedright, singlelinecheck=false}
\caption{Structure of the proposed DGCN.}
\label{bgcn}
\end{figure}

The mDAGs formed by MBDA module have special bipartite topologies, basic graph convolution operation (GCO) may lead to extensive over-smoothing problem after two GCO layers. In this paper, we propose the DGCN to apply bipartite GCO on mDAGs with residual blocks, achieving gradually contracting feature aggregation. Building on the feature matrix $\bm{X}_{in}$ and the mDAG adjacent matrix sequence $\{\bm{A}_{dag}^1,\bm{A}_{dag}^2,...,\bm{A}_{dag}^K\}$ provided by MBDA, we present the structure of DGCN in Fig.\ref{bgcn}. 

\subsubsection{Structure of DGCN}
For the series of mDAGs $\bm{A}_{dag}^k$ that have different dilation sizes, we merge these adjacent matrices into a sparse block-diagonal matrix as the input for DGCN, i.e.,
\begin{align}
    \bm{\dot{A}}_{dag}={\rm diag}(\bm{A}_{dag}^1,\bm{A}_{dag}^2,...,\bm{A}_{dag}^K)\in\mathbb{S}^{KN}.
\end{align}
Meanwhile, the feature matrix $\bm{X}_{in}$ is repeated according to the number of dilation branches $K$ to fit the shape of $\bm{\dot{A}}_{dag}$, i.e., 
\begin{align}
    \bm{\dot{X}}_{in}=[\underbrace{\bm{X}_{in}^\top,\bm{X}_{in}^\top,...,\bm{X}_{in}^\top}_K]^\top\in\mathbb{R}^{KN\times2}.
\end{align}

With the repeated feature matrix $\bm{\dot{X}}_{in}$ and the multi-level dilated adjacent matrix $\bm{\dot{A}}_{dag}$ sent into the DGCN backbone, we have the output as
\begin{align}
    \hat{\bm{\dot{X}}}_{gcn}={\rm DGCN}(\bm{\dot{X}}_{in},\bm{\dot{A}}_{dag})=[\bm{\hat{X}}_{gcn}^{1\top},\bm{\hat{X}}_{gcn}^{2\top},...,\bm{\hat{X}}_{gcn}^{K\top}]^\top,
    \label{ksolution}
\end{align}
where each $\bm{\hat{X}}_{gcn}^k=[\bm{\hat{p}}_{r_1}^k,...,\bm{\hat{p}}_{r_{N_R}}^k,\bm{\hat{p}}_{d_1}^k,...,\bm{\hat{p}}_{d_{N_D}}^k]^\top\in\mathbb{R}^{N\times 2}$ is a possible solution for (P1).
The recovery time of the $k$-th solution is calculated as
\begin{align}
    T_{rc}^k=\mathop{\rm max}\limits_{u_{r_i}\in\mathcal{U}_R}\frac{\|\bm{\hat{p}}_{r_i}^k-\bm{p}_{r_i}(t_0)\|}{v_{max}}.
\end{align}
By selecting $k^*=arg \mathop{\rm min}\limits_{k}T_{rc}^k$, we can determine the optimal target position matrix as $\bm{\hat{X}}_{gcn}^{k^*}$.

\subsubsection{Bipartite GCO}
Since the $\bm{\Dot{A}}_{dag}$ is a block-diagonal matrix, applying GCO on $\bm{\Dot{A}}_{dag}$ can be regarded as applying GCO on each $\bm{A}_{dag}^k$. Due to the special topology of $\bm{A}_{dag}^k$ as a sparse bipartite graph, we design a bipartite GCO (B-GCO) based on the Laplace matrix $\bm{L}_{dag}^k$ of the mDAG according to (\ref{degree}) and (\ref{laplace}). The features of the mDAG form a structure in non-Euclidean space, hence the features need to be mapped from the time domain into the frequency domain to achieve the convolution operation. The Fourier transform in the time domain is defined by the eigenvalue decomposition of the Laplace matrix \cite{ft}, i.e., 
\begin{align}
    \bm{L}_{dag}^k=\bm{U}\bm{\Lambda}\bm{U}^\top,
\end{align}
where the eigenvalues $\lambda_j$ denote the frequency and the eigenvectors $\bm{u}_j$ denote the Fourier modes, respectively. 

Regarding $\bm{x}$ as a signal in the mDAG, we can define the Fourier transform as $\breve{\bm{x}}={\rm FT}(\bm{x})=\bm{U}^\top\bm{x}$ and the inverse Fourier transform as $\bm{x}={\rm IFT}(\breve{\bm{x}})=\bm{U}\breve{\bm{x}}$. Therefore, the convolution between $\bm{x}$ and the convolution kernel $g$ can be represented as
\begin{align}
    g\ast\bm{x}={\rm IFT}[{\rm FT}(g)\cdot{\rm FT}(x)]=\bm{U}(\bm{U}^\top g\cdot\bm{U}^\top\bm{x})
\end{align}
where $\ast$ is the convolutional operator. Determine $g$ as a function of $\bm{L}_{dag}^k$, we can regard $\bm{U}^\top g$ as a function of the eigenvalues $g_\theta(\bm{\Lambda})={\rm diag}(\theta)$ parameterized by $\theta\in \mathbb{R}^{|\bm{x}|}$. Then, the graph convolution is transformed as
\begin{align}
    g_\theta\ast\bm{x}=\bm{U}g_\theta\bm{U}^\top\bm{x}.
    \label{ftconv}
\end{align}

The complexity of the convolution kernel in (\ref{ftconv}) can be extremely high since the eigenvalue decomposition requires $O(n^3)$ computation overhead on global information. A typical approach to decrease the computation complexity is to approximate (\ref{ftconv}) by truncated Chebyshev polynomials of the first order according to \cite{gcn}, i.e., 
\begin{equation}
\begin{aligned}
    g_\theta\ast \bm{x}
    &\approx\bm{U}(\mathop{\sum}\limits_{c=0}^1\beta_cT_c(\tilde{\bm{\Lambda}}))\bm{U}^\top\bm{x}
    =\mathop{\sum}\limits_{c=0}^1\beta_cT_c(\bm{U}\tilde{\bm{\Lambda}}\bm{U}^\top)\bm{x}\\
    &=\beta_0T_0(\tilde{\bm{L}}_{dag}^k)\bm{x}+\beta_1T_1(\tilde{\bm{L}}_{dag}^k)\bm{x}\\
    &=\beta_0\bm{x}+\beta_1(\bm{L}_{dag}^k-\bm{I}_N)\bm{x},
\end{aligned}
\end{equation}
where $T_c$ denotes the $c$-th item in the Chebyshev polynomials with $T_0(x)=1$ and $T_1(x)=x$, $\beta_c$ is the $c$-th Chebyshev coefficient, and $\tilde{\bm{\Lambda}}_d^k=\frac{2}{\lambda_{max}}\bm{\Lambda}_d^k-\bm{I}_N$ is scaled by the largest eigenvalue of $\bm{L}_d^k$. Inspired by \cite{mgc}, we define the hyper-parameter $\epsilon=-\beta_1>0$ and let $\beta_0 = \beta_1+1$, then we have the B-GCO of signal $x$ in the mDAG as
\begin{align}
    g_\theta\ast \bm{x}=(\bm{I}_N-\epsilon\bm{L}_{dag}^k)\bm{x}.
\end{align}

Subsequently, we extend the $g_\theta$ into a trainable B-GCO layer. In each B-GCO layer, we introduce a weight matrix for feature linear transformation, which serves as the trainable parameter. Specifically, in the $q$-th B-GCO layer, the feature matrix is processed as
\begin{align}
    \bm{\dot{X}}_{gco}^q={\rm GCO}(\bm{\dot{X}}_{gco}^{q-1},\bm{\dot{A}}_{dag})
    =(\bm{I}_{KN}-\epsilon\bm{\dot{L}}_{dag})\bm{\dot{X}}_{gco}^{q-1}\bm{W}^q,
    \label{w}
\end{align}
where the Laplace matrix $\bm{\dot{L}}_{dag}$ is computed based on $\bm{\dot{A}}_{dag}$, and $\bm{W}^q$ is a layer-specific trainable weight matrix.

\subsubsection{DGCN backbone}

As shown in the right part of Fig.\ref{bgcn}, DGCN backbone comprises an input normalization layer, the bipartite GCO layers, the nonlinear activation layers, and an output upscale layer.

The position normalization layer first down-scales the input features in$\bm{\Dot{X}}_{in}$ to the range of $(-1,1)$. Let $\bm{p}_c=\frac{1}{N}\sum_{i=1}^N\bm{p}_{i}(t_0)$ denote the center point of the USNET at time $t_0$, hence we can define the scale ratio $D=\mathop{\rm{max}}\limits_{u_i\in\mathcal{U}}\|\bm{p}_c-\bm{p}_i(t_0)\|$ is the maximum distance between nodes and $\bm{p}_c$. The normalized feature matrix as be calculated as
\begin{align}
    \bm{\Dot{X}}_{norm}=\frac{\bm{\Dot{X}}_{in}-\bm{p}_c^\top}{D+1}.
\end{align}

Next, the normalized feature matrix is sent into a single B-GCO layer, additionally with the ReLU activation function as the non-linear layer. The output feature matrix is 
\begin{align}
    \bm{\dot{X}}_{gco}^1={\rm ReLU}\left({\rm GCO}(\bm{\dot{X}}_{norm},\bm{\dot{A}}_{dag})\right).
\end{align}
Here, the feature dimension is extended from 2 to the hidden dimension $d_s$ by setting $\bm{W}^1$ as a 2-by-$d_s$ weight matrix.

Subsequently, the output feature matrix is processed by $L$ residual blocks. Each residual block consists of two B-GCO layers and two ReLU layers, plus a residual connection with $\bm{\dot{X}}_{gco}^1$ at the output. Therefore, the output feature matrix of the $l$-th residual block is
\begin{align}
    \bm{\dot{X}}_{block}^l=\left[{\rm ReLU}\left({\rm GCO}(\bm{\dot{X}}_{block}^{l-1},\bm{\dot{A}}_{dag})\right)\right]^2+\bm{\dot{X}}_{gco}^1.
\end{align}
The weight matrices in the residual block are all set as $d_s$-by-$d_s$ matrices. Particularly, we have $\bm{\dot{X}}_{gco}^1$ as the input of the first residual block. Note that dropouts \cite{dropout} can be added between two residual blocks
to increase the ability of generalization.

After $L$ residual blocks, the final B-GCO layer brings the feature dimension back to 2 with the weight matrix set as a $d_s\times2$ dimensional matrix. In addition, we introduce a tanh function to map the feature into the range of $(-1,1)$. Therefore, the DGCN has a total of $Q=2L+2$ GCO layers with the output
\begin{align}
    \bm{\dot{X}}_{gco}^Q={\rm tanh}\left({\rm GCO}(\bm{\dot{X}}_{block}^L,\bm{\dot{A}}_{dag})\right).
\end{align}

Finally, the output layer up-scales the feature matrix to the target position ranges. The final output position matrix of DGCN is calculated as
\begin{align}
    \hat{\bm{\Dot{X}}}_{gcn}=(D+1)\cdot(\bm{\dot{X}}_{gco}^Q+\bm{p}_c^\top).
\end{align}

\subsubsection{Loss function design}
Denote the loss function of the DGCN as $\mathcal{L}(\bm{\dot{X}}_{in},\mathcal{W})$, where $\bm{\dot{X}}_{in}$ is the input feature matrix and $\mathcal{W}=\{\bm{W}^1,\bm{W}^2,...,\bm{W}^Q\}$ is the set of layer weight matrices. 

Note that the output of DGCN $\hat{\bm{\Dot{X}}}_{gcn}$ contains $K$ possible solutions according to (\ref{ksolution}). Hence, we can calculate a loss $\mathcal{L}^k$ under each solution $\bm{\hat{X}}_{gcn}^k$ respectively, where the design of $\mathcal{L}^k$ should be consistent with (P1).
Specifically, we rewrite the problem (P1) as
\begin{align}
    ({\rm P}1^\dagger):\quad \mathop{\rm min}\limits_{\bm{\hat{X}}_{gcn}^k}\quad & T_{rc}^k=\mathop{\rm max}\limits_{u_{r_i}\in\mathcal{U}_R}\frac{\|\bm{\hat{p}}_{r_i}^k-\bm{p}_{r_i}(t_0)\|}{v_{max}}
    \label{p2s}\\
    \quad {\rm s.t.}\quad & \hat{N}_S^k - 1 \leq 0,
\end{align}
where $T_{rc}$ is substituted by the recovery time $T_{rc}^k$ of the solution $\bm{\hat{X}}_{gcn}^k$, and $\hat{N}_S^k$ is the number of sub-nets in the remaining graph recovered by $\bm{\hat{X}}_{gcn}^k$. The equality constraint (\ref{c2}) is represented by the inequality $\hat{N}_S^k - 1 \leq 0$. Therefore, we can form a basic loss function of the $k$-th solution as
\begin{align}
    \mathcal{L}^k=T_{rc}^k+\lambda (\hat{N}_S^k - 1),
    \label{loss1}
\end{align}
where the Lagrange multiplier $\lambda$ is a positive constant.
Based on (\ref{loss1}), we define the total training loss of the DGCN as
\begin{align}
    \mathcal{L}_{total}=\mathop{\sum}\limits_{k=1}^K\mathcal{L}^k=\mathop{\sum}\limits_{k=1}^KT_{rc}^k+\lambda (\hat{N}_S^k - 1).
\end{align}


\subsection{Theoretical Analysis of ML-DAGL}
To analyze ML-DAGL, we need to address whether the solution can guarantee connectivity recovery and assess its performance under worst-case scenarios. We first demonstrate that the B-GCO always converges to a feasible solution on a connected mDAG, and derive the selection of hyper-parameters to form a connected mDAG. Subsequently, we analyze the theoretical upper bound for the recovery time.

\subsubsection{Convergence of B-GCO}
Since $\bm{\dot{A}}_{dag}$ is block-diagonal, we can regard the B-GCO on feature matrix $\bm{\dot{X}}_{in}$ as applying B-GCO on $\bm{X}_{in}$ for $K$ times with different $\bm{A}_{dag}^k$. For each B-GCO with $\bm{A}_{dag}^k$, we can prove that it is a contraction operation for each iteration as follows. 

\begin{proposition}
In the metric space of position matrices $\{\bm{X}_{gco}\}\subseteq\mathbb{R}^{N\times 2}$, the B-GCO with $\bm{A}_{dag}^k$ is a contraction operation when $0<\epsilon\leq\frac{1}{\|\bm{A}_{dag}^k\|_\infty}$, and there exists and only exists one position matrix $\bm{\bar{X}}_{gco}$ such that
\begin{align}
    \mathop{\rm lim}\limits_{q\rightarrow\infty}(\bm{I}_N-\epsilon\bm{L}_{dag}^k)^q\bm{X}_{gco}=\bm{\bar{X}}_{gco}.
\end{align}
Especially, when B-GCO takes $\bm{X}_{in}$ as the input, $\bm{\bar{X}}_{gco}\equiv[\bm{p}_c,\bm{p}_c,...,\bm{p}_c]^\top$ holds when the $k$-hop mDAG is connected.
\label{prop-c2}
\end{proposition}

\begin{IEEEproof}
See Appendix A.
\end{IEEEproof}

Proposition \ref{prop-c2} demonstrates that, after a sufficient number of iterations, the output matrix $\bm{\hat{X}}_{gco}^k$ of a connected mDAG will converge to $\bm{\bar{X}}_{gco}^k$. Therefore, B-GCO can always provide a feasible solution for (P1) since the remaining graph constructed from $\bm{\bar{X}}_{gco}$ is fully connected. 
To guarantee the convergence of B-GCO on varying damage cases, we need to find a proper $K$, the number of parallel branches, to ensure that at least one mDAG is connected, and determine a proper $\epsilon$ to meet the convergence constraint.

\subsubsection{Choice of hyper-parameters}
As the scales of nodes $N$ and damage $N_D$ increase, mDAG with a larger dilation size $k$ is more likely to be connected. It can be easily observed that once a $k$-hop mDAG with $\bm{A}_{dag}^k$ is connected, then for every $k<k_i\leq K$, the $k_i$-hop mDAG is connected. Note that $k$ is limited by $k\leq K$, therefore, we only need to ensure that the mDAG on the $K$-th branch is always connected in the worst-case scenario. 
Denote $H_{max}$ as the maximum number of hops in the original USNET, we set the choice of $K$ as
\begin{equation}
    K = \left\lfloor \frac{H_{max}+1}{2} \right\rfloor.
    \label{K}
\end{equation}
where $\lfloor\cdot\rfloor$ denotes the floor function.

For the choice of $\epsilon$, we have $\|\bm{A}_{dag}^k\|_\infty$ positively correlated to the dilation size $k$. Note that constraint $\|\bm{A}_{dag}^k\|_\infty<N$ always holds. To ensure that the convolution kernel $g_\theta$ remains isomorphic for different $k$ during the batch process, we set $\epsilon=\frac{1}{N}<\frac{1}{\|\bm{A}_{dag}^k\|_\infty}$ for the bipartite graph convolution to meet the convergence constraint. With the $K$ in (\ref{K}), the ML-DAGL algorithm must provide a solution that guarantees connectivity restoration, hence the convergence holds.

\subsubsection{Upper bound of recovery time}
Since the final solution $\bm{\hat{X}}_{gcn}^{k^*}$ will approach $\bm{\bar{X}}_{gcn}$ when the mDAG is connected, the solution under the worst case refers to $\bm{\hat{X}}_R=\bm{\bar{X}}_{gco}=[\bm{p}_c,...,\bm{p}_c]^\top$ that guarantees the restoration of connectivity. Therefore, we can give an upper bound for the recovery time as
\begin{align}
    T_{rc}^{max}=sup\{T_{rc}^{k^*}\} = \frac{\mathop{\rm max}\limits_{u_{r_j}\in \mathcal{U}_R}\|\bm{p}_c-\bm{p}_{r_j}(t_0)\|}{v_{max}}.
\end{align}

In practical applications, the model often finds a more efficient solution based on early termination before it converges to $\bm{\bar{X}}_{gco}$. According to Appendix A, when the number of iterations is small, the remaining UAV tends to aggregate around their neighboring damaged region. This characteristic contributes to the performance advantage of the ML-DAGL algorithm.

\subsection{Computational complexity analysis}
The computational complexity of the proposed ML-DAGL consists of the complexities of both MBDA module and DGCN. Since the structure of the ML-DAGL has $K$ dilation branches, we first analyze the computational complexity of the 1-hop branch, and then generalize it to the $k$-hop branch. 

\subsubsection{Complexity of MBDA}
Matrix operations in MBDA module is the Hadamard product in (\ref{Hadamard}). 
Although MBDA enhance the sparsity of $\bm{A}_{in}^k$, the damage attention matrix $\bm{M}_{da}$ is always a sparse matrix, hence the computational complexity of the k-hop branch is calculated as $\mathcal{O}\left(\text{nnz}(\bm{A}_{in}^k)\cdot\text{nnz}(\bm{M}_{da})\right)$, where $\text{nnz}(\cdot)$ is the number of non-zero elements in a matrix. 

For the 1-hop branch, assuming that $N$ nodes are randomly distributed within a $D\times D$ square area, two nodes have a probability of $\frac{\pi d_{tr}^2}{D^2}$ to become neighbors, hence the node degree follows a Poisson distribution. The number of non-zero elements in $\bm{A}_{in}^1$ can be calculated as
\begin{align}
    \text{nnz}(\bm{A}_{in}^1)=\bar{d}N=\pi d_{tr}^2\rho N,
\end{align}
where $\bar{d}$ is the average node degree, $d_{tr}$ is the maximum communication range determined by (\ref{dtr}), $\rho=N/D^2$ is the node density. Meanwhile, we have
\begin{align}
    \text{nnz}(\bm{M}_{da})=2N_DN_R=2p(1\!-\!p)N^2,
\end{align}
where $p=N_D/N$ denotes the damage ratio. Therefore, for the 1-hop branch of the MBDA module, the entire complexity is $\mathcal{O}\left(2p(1\!-\!p)\bar{d}N^3\right)$.

For the $k$-hop branch, we can regard the $k$-hop dilation as a node extends its communication range $k$ times, hence the node degree in $\bm{A}_{in}^k$ still follows the Poisson distribution, hence we have $\text{nnz}(\bm{A}_{in}^k)=k^2\bar{d}N$. The computational complexity of the MBDA module is finally determined as $\mathcal{O}\left(\frac{1}{3}p(1\!-\!p)K(K\!-\!1)(2K\!-\!1)\bar{d}N^3\right)$. 

Consequently, the MBDA module requires a computational complexity of $\mathcal{O}\left(K^3N^3\right)$. Since the MBDA module is a data pre-processing stage, it runs only once for each CNS issue with necessary complexity.

\subsubsection{Complexity of DGCN}
For DGCN, the matrix operations are mainly matrix multiplications in B-GCO layers. The total complexity is related to the number of B-GCO layers and the round of iterations. Therefore, we first determine the complexity of DGCN within one iteration.
\begin{proposition}
The one-round computational complexity of the DGCN is given by
\begin{align}
\mathcal{O}\left(KN\mathop{\sum}\limits_{q=1}^Q[p(1\!-\!p)\pi d_{tr}^2\rho(K\!-\!1)\!+\!1\!+\!d_{out}^q]d_{in}^q\right),
\end{align}
where $Q$ is the number of B-GCO layers, $d_{in}^q, d_{out}^q$ are the input and output dimensions of the $q$-th layer.
\label{prop-cp}
\end{proposition}

\begin{IEEEproof}
See Appendix B.
\end{IEEEproof}

Therefore, the DGCN has a linear computational complexity of $\mathcal{O}\left(K^2N\right)$ to $N$ for a single iteration. This is because the sparsity of the B-GCO kernel provides a significant reduction in the complexity of matrix multiplication during graph convolutions, according to Proposition~\ref{prop-cp}. 

In summary, the proposed algorithm has total computational complexity of $\mathcal{O}\left(K^3N^3+RK^2N\right)$, where $R$ is the round of online iterations. With the pre-trained model, ML-DAGL algorithm requires a small $R$ to compute the optimal solution, hence the computational complexity is acceptable.

\section{Simulation Results}
This section first introduces the setup of simulation experiments and then presents the performance of the proposed algorithm \footnote{The source codes of the proposed ML-DAGL algorithm are available on \textit{https://github.com/lytxzt/Damage-Attentive-Graph-Learning}. The computations in this paper were run on the $\pi$ 2.0 cluster supported by the Center for High Performance Computing at Shanghai Jiao Tong University.}. The performances of center-fly, HERO \cite{hero}, SIDR \cite{sidr}), and the graph-learning algorithms including original GCN \cite{gcn}, GAT\cite{gat}, CR-MGC \cite{mgc}, DEMD \cite{demd}, are also displayed for comparisons. 

\subsection{Simulation Setup}
\subsubsection{Environment initialization}
In the simulation, the original USNET is composed of $N$ UAVs randomly deployed within a square area of size $D \times D$, with a fixed node density of $\rho = 200\,\text{nodes/km}^2$. A fraction of the UAVs is assumed to be destroyed due to attacks or failures, where the number of failed UAVs is given by $N_D = pN$, with $p$ representing the damage ratio. The maximum allowed recovery time is set to be the time cost of flying across the half area, i.e., $T_{rc}^{max} = D / (2v_{max})$. Solutions fail to restore connectivity within $T_{rc}^{max}$ are regard as not convergent.

For constant parameters, the communication radius is set to $d_{tr} = 120\,\text{m}$ following the analysis in \cite{mgc}. The maximum UAV velocity is constrained to $v_{max} = 10\,\text{m/s}$ for computational convenience. Recovery is modeled in discrete time steps with a fixed interval of $\Delta t = 0.1\,\text{s/step}$. 

\subsubsection{Model Parameters}
We develop $L=3$ residual blocks in DGCN and set the hidden feature dimension to $d_s=512$. The Adam optimizer with a learning rate of 0.0001 is applied for training and the dropout rate is set to 0.1 to prevent overfitting. The pre-trained models are trained for 500 iterations under randomly $p = 0.5$ damage scenarios with different swarm scale $N=20,50,100,200,500,1000$. 

\subsubsection{Evaluation Metrics}
Three metrics are adopted to evaluate the performance of resilience. The \textit{convergent Ratio} $R_c$ denotes the proportion of simulations that can restore connectivity within the allowed time $T_{rc}^{max}$, which reflects the robustness of algorithms. The \textit{recovery Time} $T_{rc}$ is the average time required to complete the connectivity restoration process, offering a direct measure of the algorithm’s efficiency. The \textit{degree Distribution} $P_d$ denotes the proportion of nodes whose degrees do not exceed $d$ in the recovered network. The mean degree $E(d)$ and maximum degree $max(d)$ also represent the distribution characteristics.

\begin{figure*}[!t]
\centering{
\subfloat[convergent ratio under varying $N_D$]{\includegraphics[width=.48\linewidth, height=.37\linewidth]{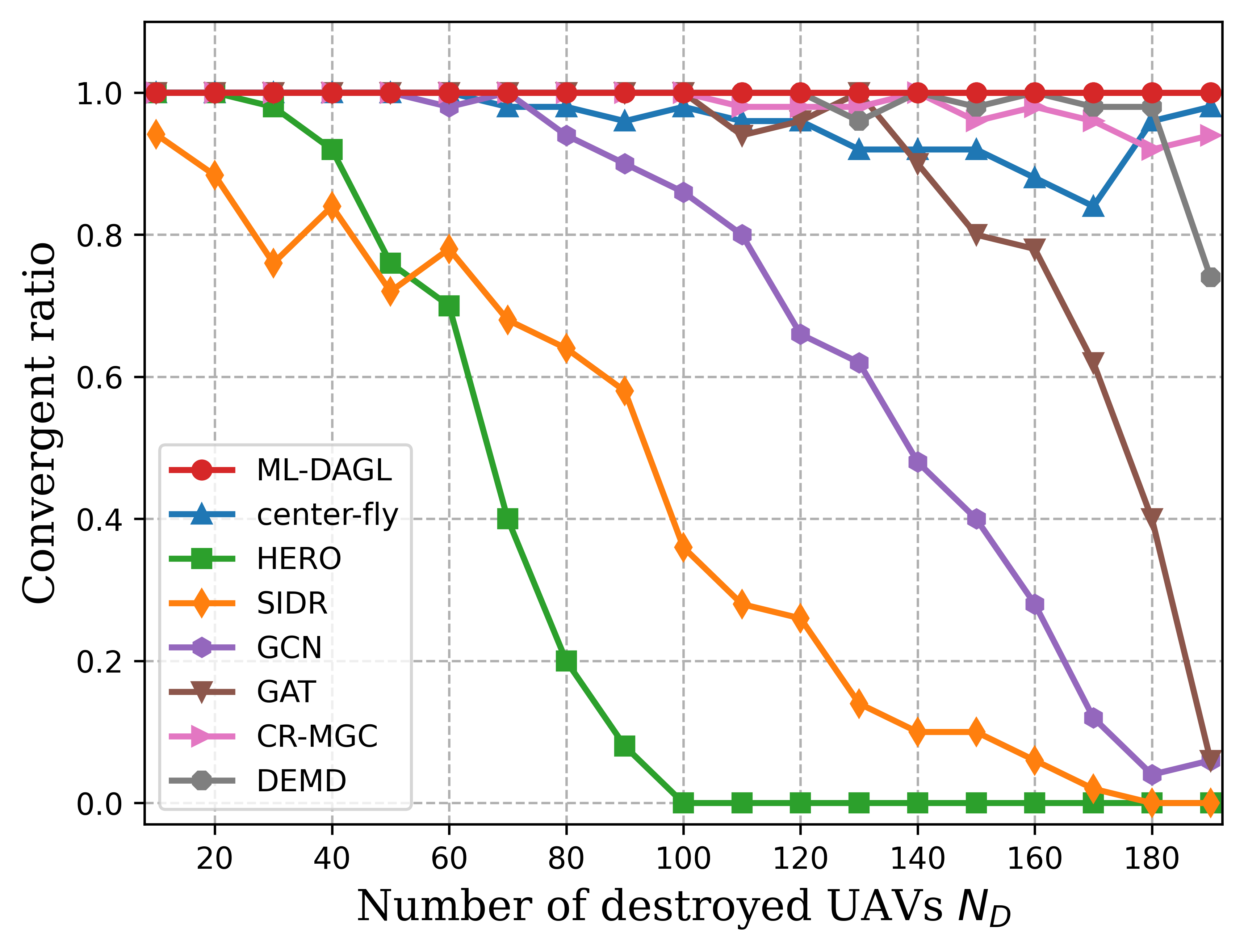}\label{res1}}
\subfloat[average recovery time under varying $N_D$]{\includegraphics[width=.48\linewidth, height=.37\linewidth]{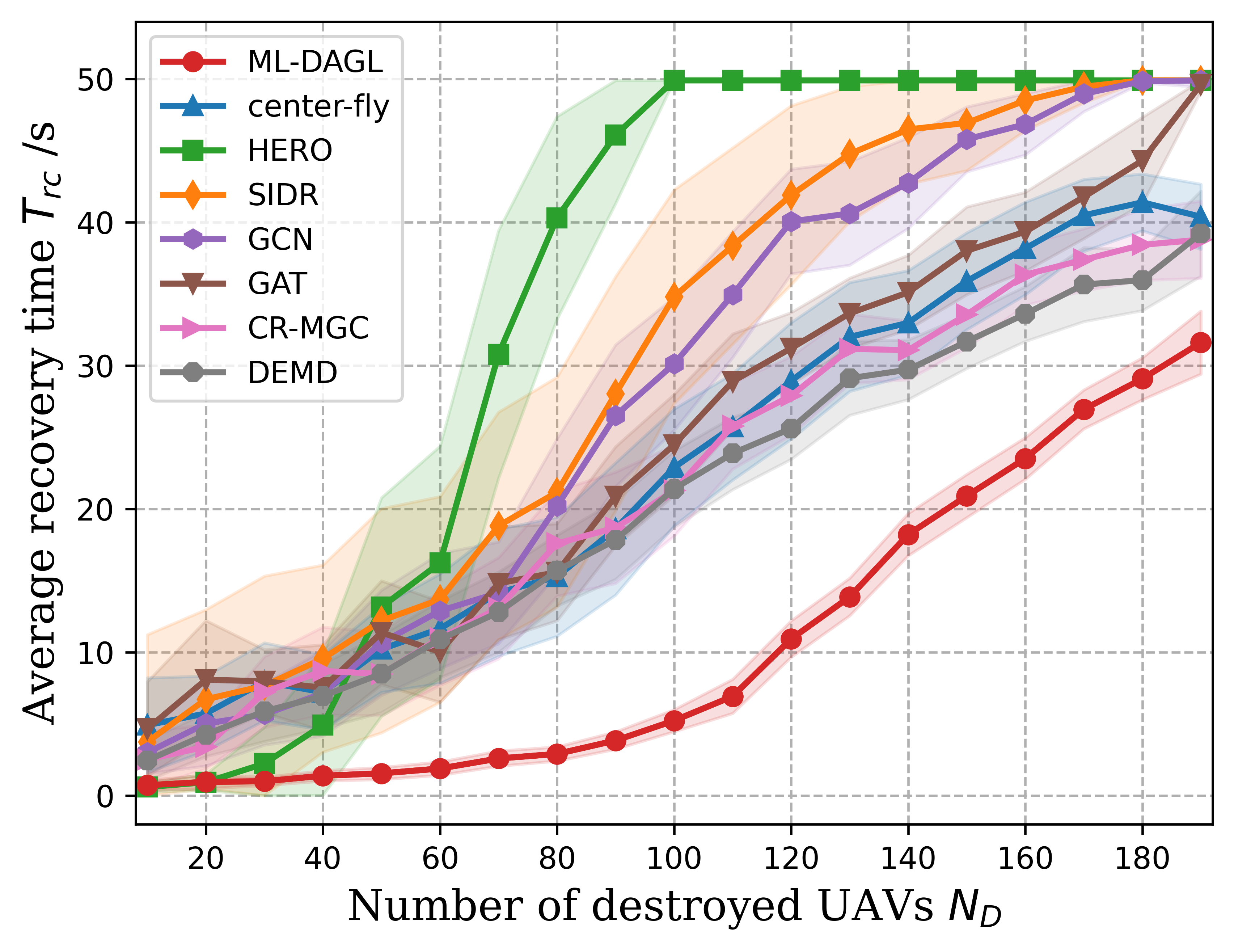}\label{res2}}\\
\subfloat[cumulative degree distributions under $N_D=100$]{\includegraphics[width=.48\linewidth, height=.37\linewidth]{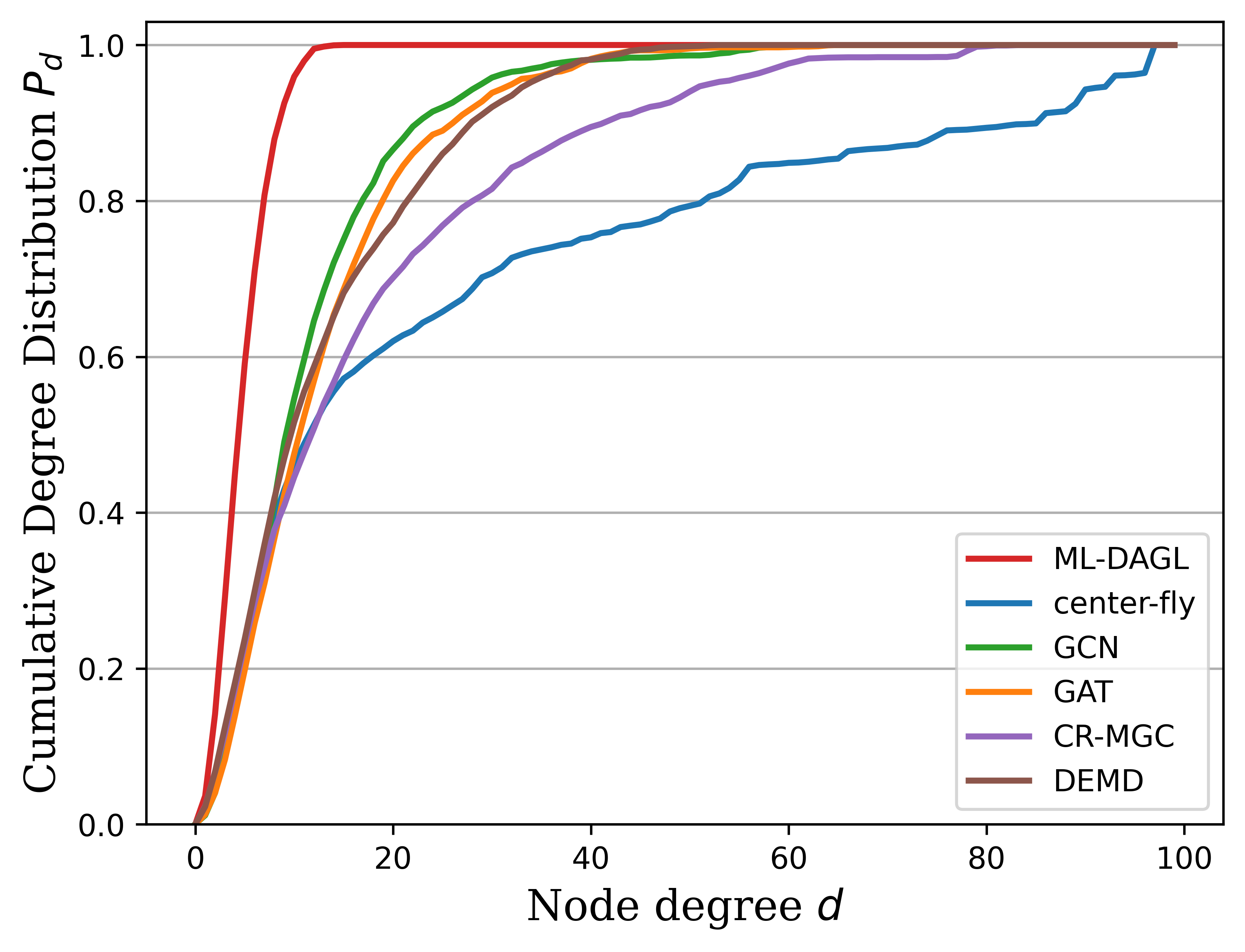}\label{res3}}
\subfloat[cumulative degree distributions under $N_D=150$]{\includegraphics[width=.48\linewidth, height=.37\linewidth]{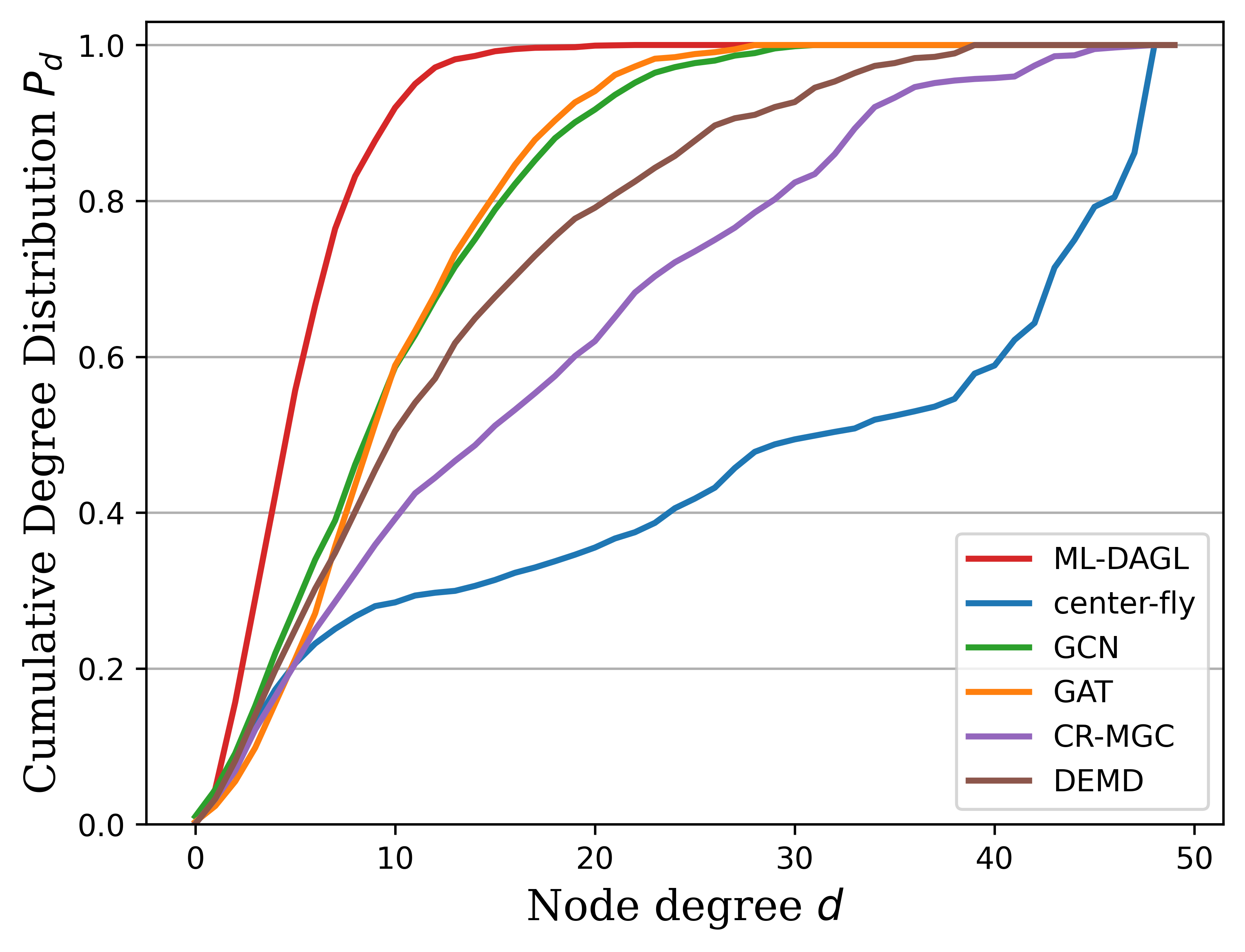}\label{res4}}}
\captionsetup{justification=raggedright, singlelinecheck=false}
\caption{Statistical results of resilient algorithms under varying damage scenarios with swarm scale $N=200$.}
\label{adaptability}
\end{figure*}

\subsection{Statistical Results Evaluation}
We develop $N=200$ UAVs within a $1\,\text{km} \times 1\,\text{km}$ area and randomly destroy between 10 and 190 UAVs to evaluate the performance of the proposed algorithm under varying damage levels. Therefore, The period of recovery process is $T_{rc}^{max} = 50\,\text{s}$. Each scenario is repeated 50 times for statistical robustness. Simulation results for resilient algorithms are presented in Fig.\ref{adaptability}.

The convergent ratios $R_c$ of resilient algorithms are illustrated in Fig.\ref{res1}, showing that $R_c$ for HERO, SIDR, GCN, and GAT declines sharply with increasing $N_D$, while center-fly, CR-MGC, and DEMD maintain near-perfect convergence up to $N_D = 150$. In contrast, the proposed ML-DAGL consistently achieves $R_c = 1$, successfully resolving all connectivity restoration cases, including under extreme damage scenarios.

The average recovery time $T_{rc}$ verses $N_D$ are plotted in Fig.\ref{res2}. It can be observed that all approaches require longer $T_{rc}$ when $N_D$ increases. The curve of ML-DAGL algorithms first grows slowly then increase rapidly after $N_D$ reaching 100. On the other hand, ML-DAGL significantly outperforms other approaches, reducing average recovery time by 59.8\% compared with second-best approach. For instance, the proposed algorithm achieves 81.7\%, 71.0\%, and 30.0\% reductions compared to DEMD under $N_D = 50$, 100, and 150, respectively. These results confirm ML-DAGL's efficiency across varying damage scales.

The cumulative node-degree distributions $P(d)$ for $N_D = 100$ and $150$ are shown in Fig.\ref{res3} and Fig.\ref{res4}. The degree of recovered network by ML-DAGL is distributed in lower region, while other approaches have more high-degree nodes. This demonstrates that the proposed algorithm mitigates the over-aggregation and thereby forms more even topologies.

\subsection{Case Study}

\begin{figure*}[!t]
\vspace{-1em}
\centering{
\subfloat[the remaining graph with 7 sub-nets]{\includegraphics[width=.48\linewidth, height=.37\linewidth]{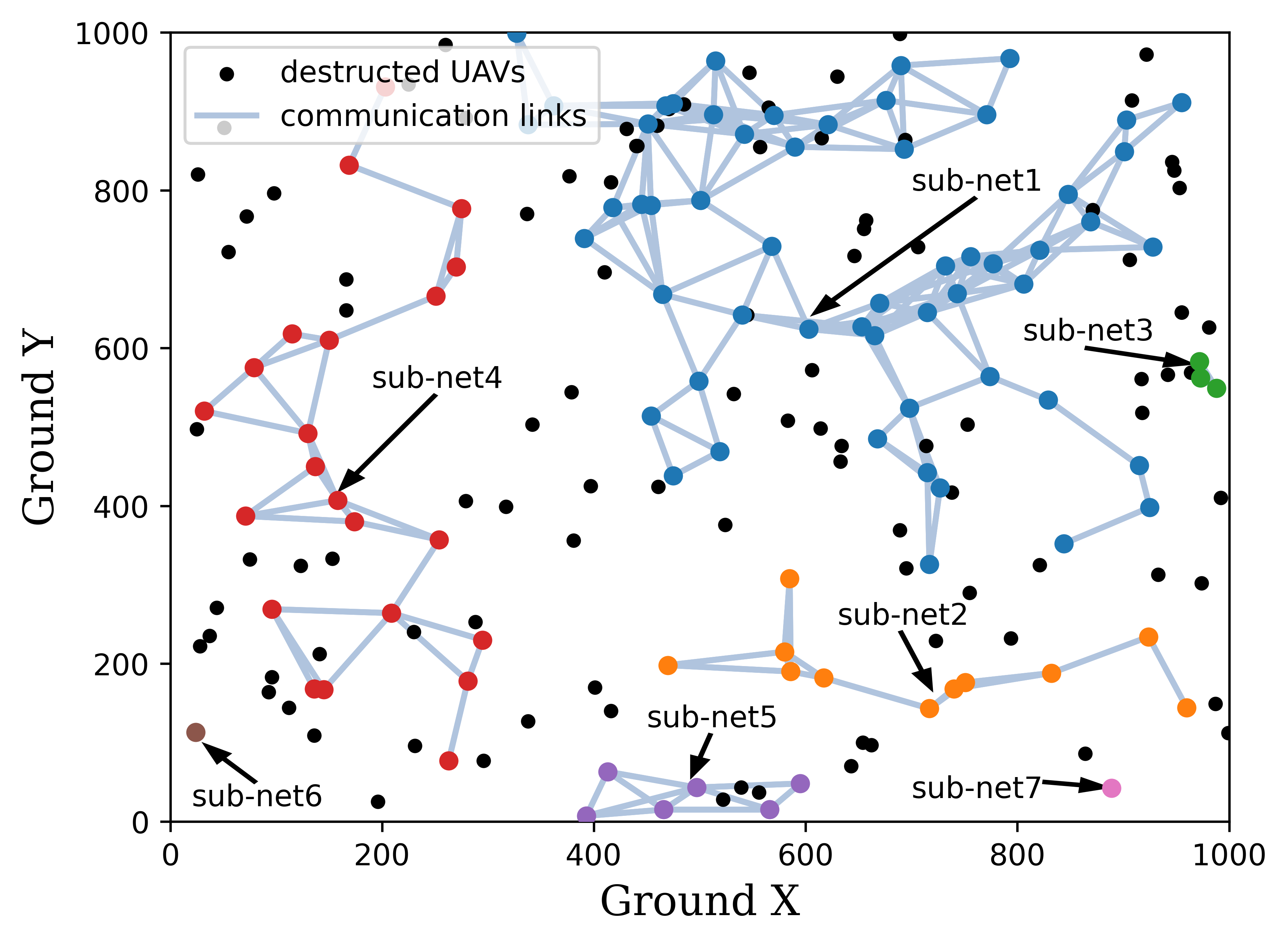}\label{case1}}
\subfloat[the number of sub-nets during recovery versus time]{\includegraphics[width=.48\linewidth, height=.37\linewidth]{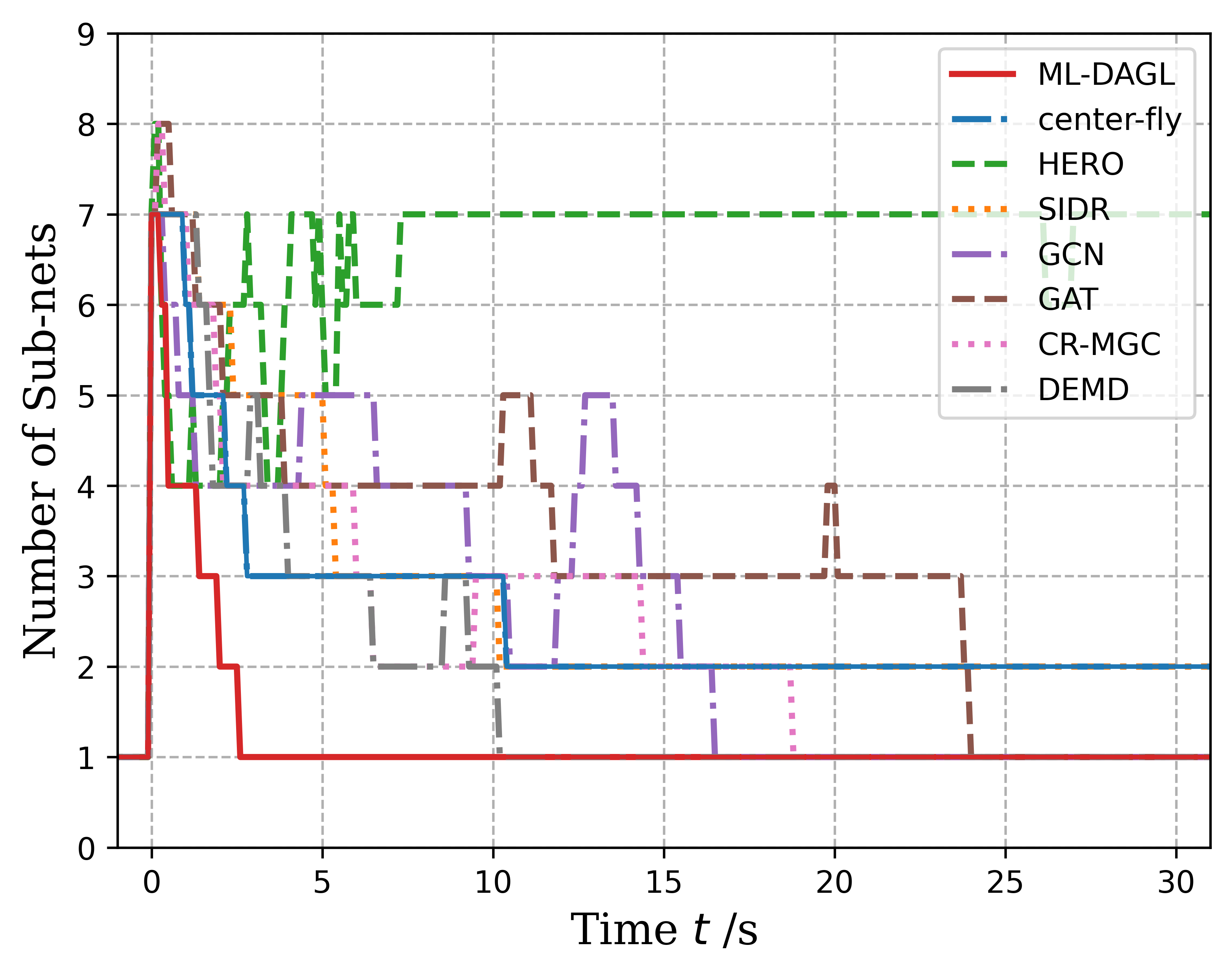}\label{case2}}\\
\subfloat[recovery trajectories of CR-MGC]{\includegraphics[width=.33\linewidth]{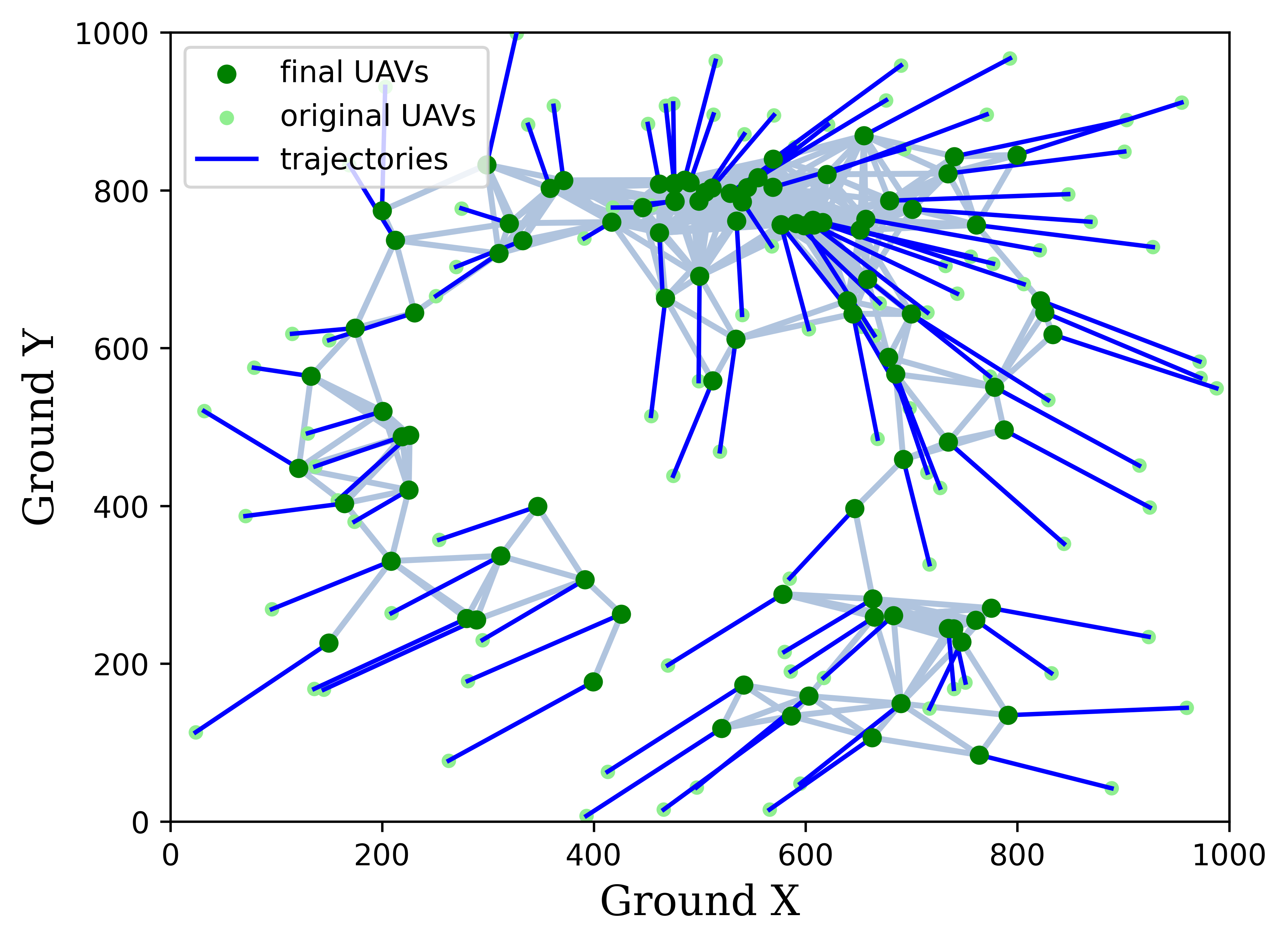}\label{case3}}
\subfloat[recovery trajectories of DEMD]{\includegraphics[width=.33\linewidth]{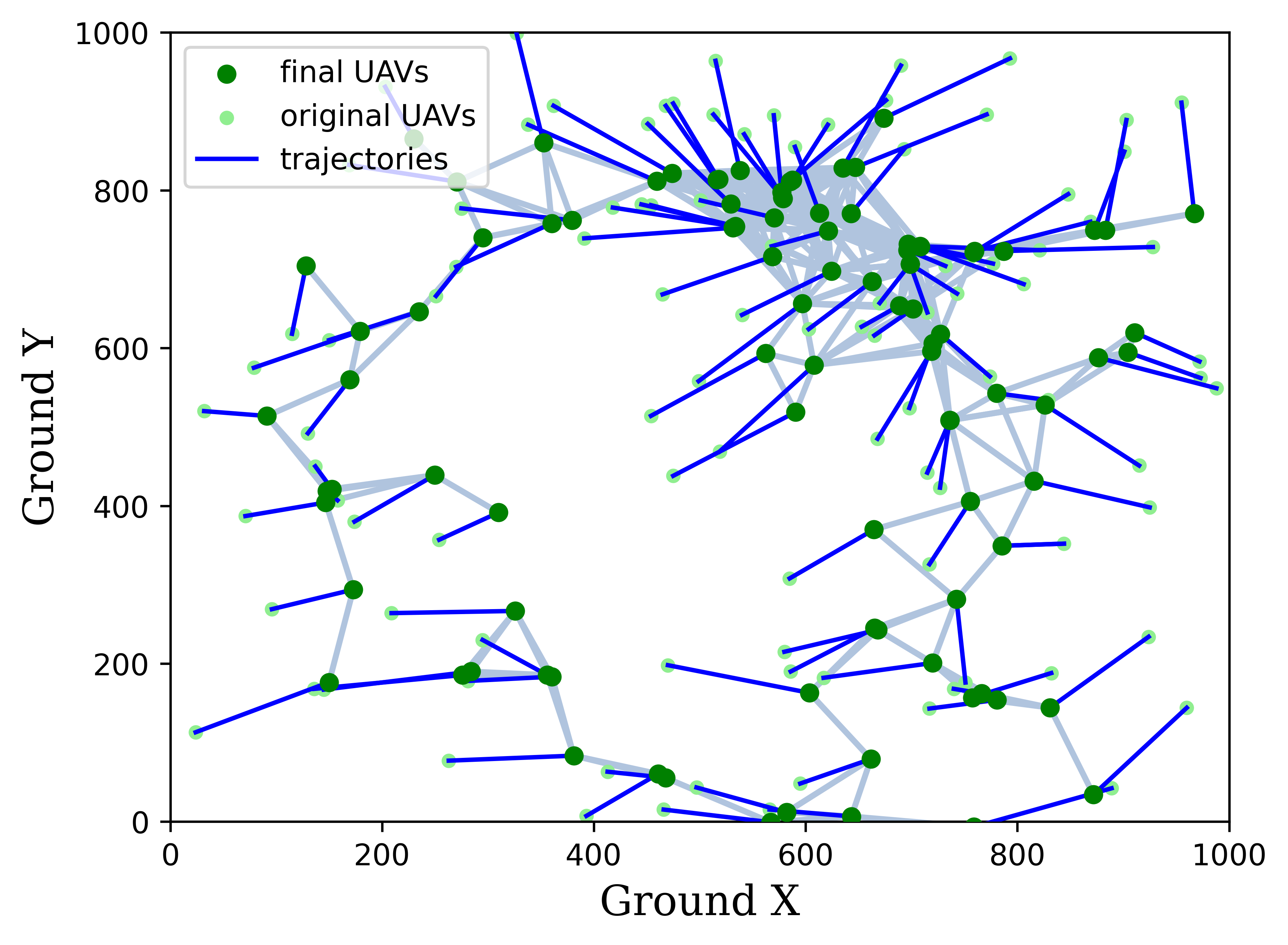}\label{case4}}
\subfloat[recovery trajectories of ML-DAGL]{\includegraphics[width=.33\linewidth]{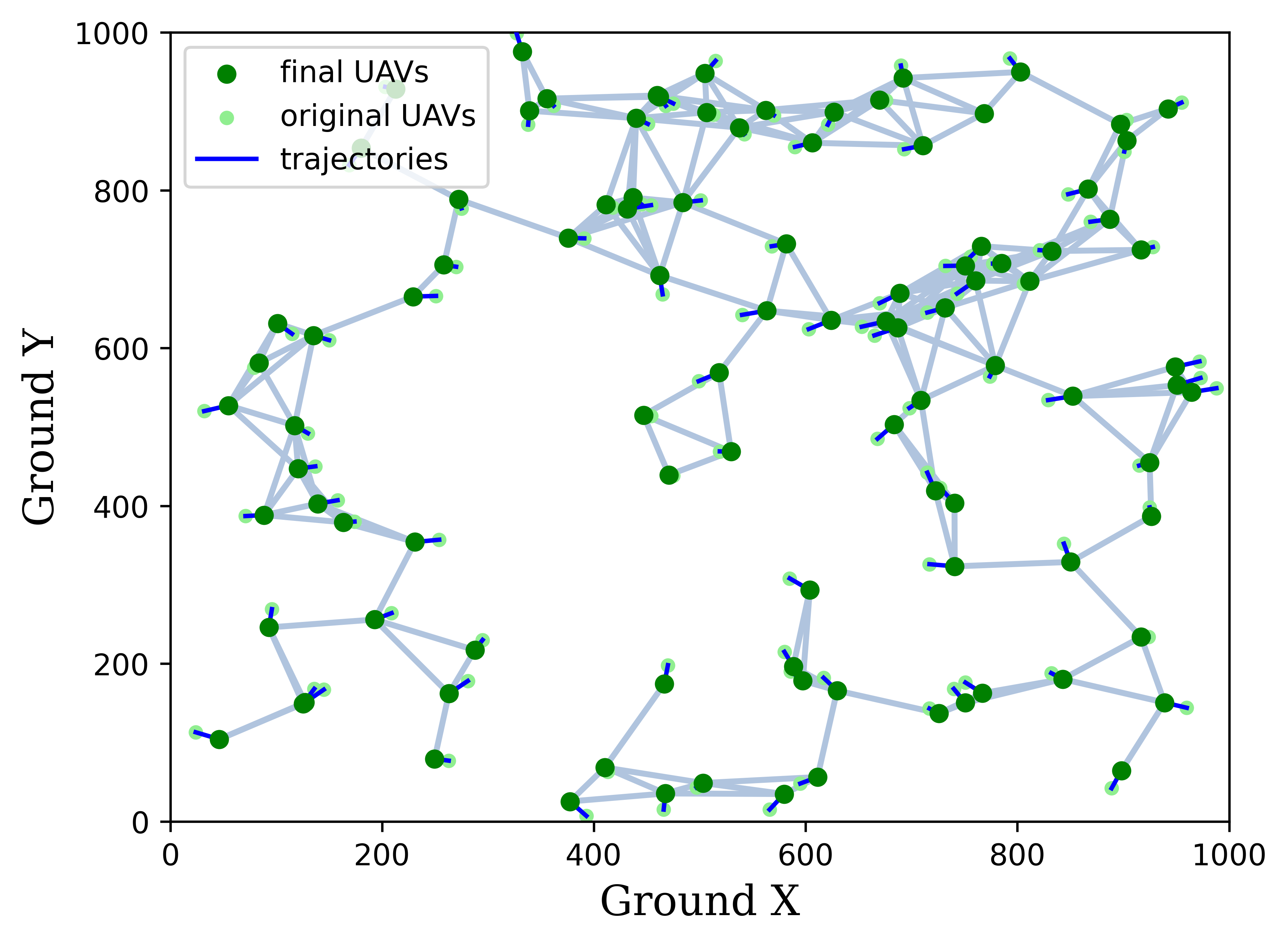}\label{case5}}}
\captionsetup{justification=raggedright, singlelinecheck=false}
\caption{Results of the case study of different algorithms.}
\label{case}
\end{figure*}

To provide a more intuitive comparison of the recovery performance across different resilient algorithms, we present an example illustrating the network restoration process under each approach as the case study, as shown in Fig.\ref{case}.

In Fig.\ref{case1}, $N_D = 100$ UAVs are randomly removed from the original USNET, resulting in a split topology with 7 disconnected sub-nets of different sizes. The numbers of sub-nets during recovery are plotted in Fig.\ref{case2}. The proposed ML-DAGL algorithm restores connectivity in just 2.4 seconds, whereas other approaches require substantially more time.

The recovery trajectories produced by CR-MGC, DEMD, and ML-DAGL are shown in Fig.\ref{case3} to Fig.\ref{case5}, respectively. We can see that the trajectories of CR-MGC and DEMD tend to converge toward sub-net 1, which has the largest size and highest degrees for aggregation. In contrast, the trajectories under ML-DAGL are notably shorter, with UAVs moving toward nearby damaged regions rather than dominant clusters. This leads to more distributed solutions and uniform topologies, thereby mitigating local over-aggregation.

\subsection{Online Iteration Consumption}
\begin{figure}[!t]
\vspace{-1em}
\centerline{\includegraphics[width=.98\linewidth, height=.74\linewidth]{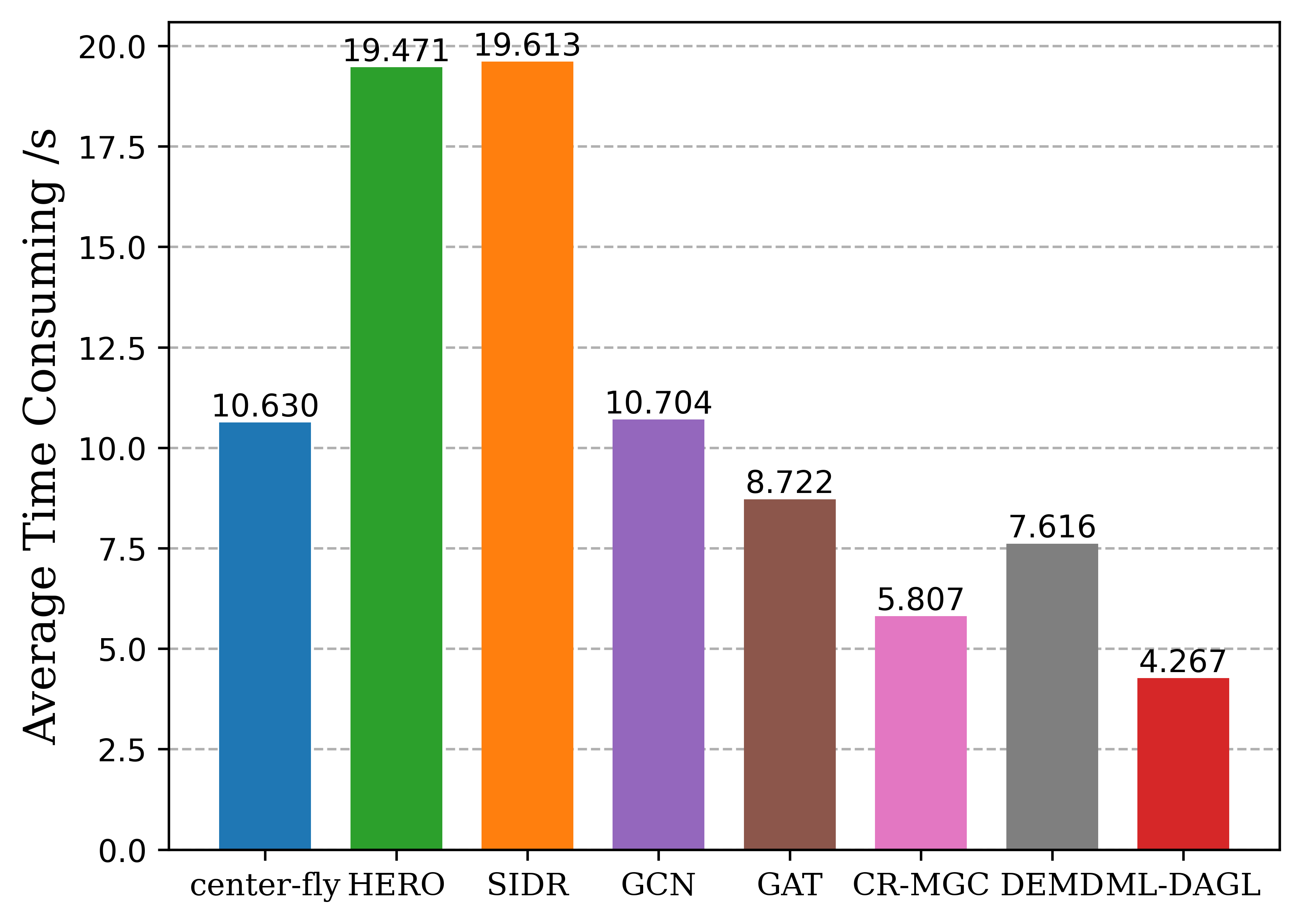}}
\caption{Results of total online average time consumption of different algorithms under $N=200$.}
\label{time}
\end{figure}

The average time consumptions of online iteration with different algorithms are presented in Fig.\ref{time}. Note that center-fly, HERO, and SIDR perform online computations at each execution step, their total time consumptions get pretty high under pro-longed recovery processes. In contrast, graph learning-based algorithms compute the recovery strategy only once upon, and the proposed ML-DAGL demonstrates competitive time cost for solving CNS issues. For practice, ML-DAGL needs only 6.02 Mb parameters and 513 MiB running memory for USNET with 200 UAVs, showing acceptable complexity for practical deployment.

\subsection{Ablation Study}
To analyze the effectiveness of each component in the ML-DAGL algorithm framework, we conduct several ablation experiments under the swarm scale set to $N=200$. 

\subsubsection{Study of pre-trained model}
\begin{figure}[!t]
\centerline{\includegraphics[width=.98\linewidth, height=.74\linewidth]{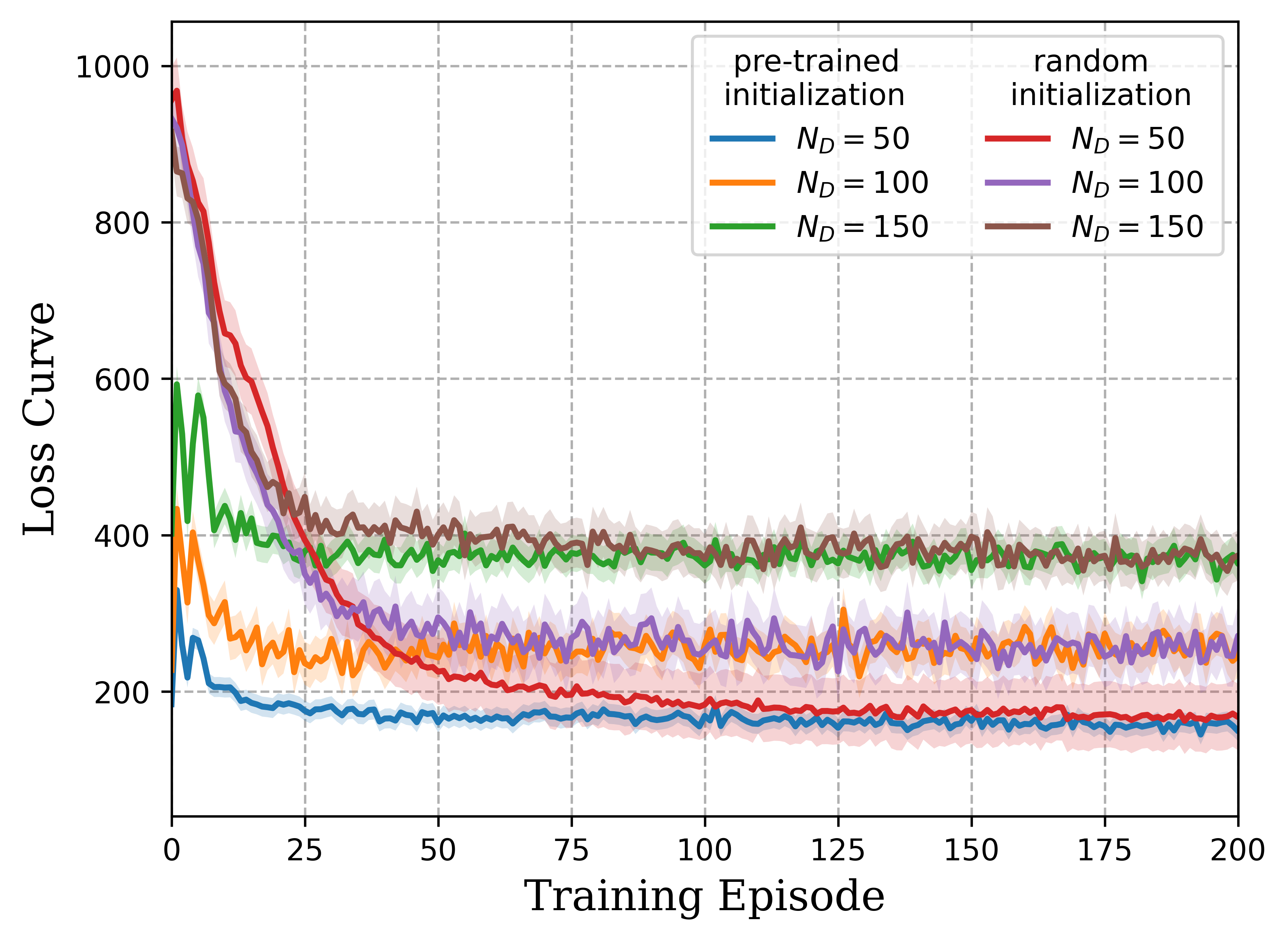}}
\caption{Loss curve of ML-DAGL training with pre-trained and random initialization under different $N_D$.}
\label{loss}
\end{figure}

We represent the loss curves of ML-DAGL during training in Fig.\ref{loss}, where the weights $W$ in (\ref{w}) are initialized using either a pre-trained model or random parameters. The results show that the curves with the pre-trained weights begin with lower loss values and reach a stable phase at the 50th episode. In contrast, curves of random initialization require more iterations to converge according to $N_D$. Furthermore, we can see that the same pre-trained model works regardless of variations in $N_D$, indicating that only a single pre-trained model for a given $N$ is necessary. The implementation of the pre-trained model can significantly reduce the training time with limited storage requirements.


\subsubsection{Study of MBDA}
\begin{figure}[!t]
\vspace{-1em}
\centerline{\includegraphics[width=.98\linewidth, height=.74\linewidth]{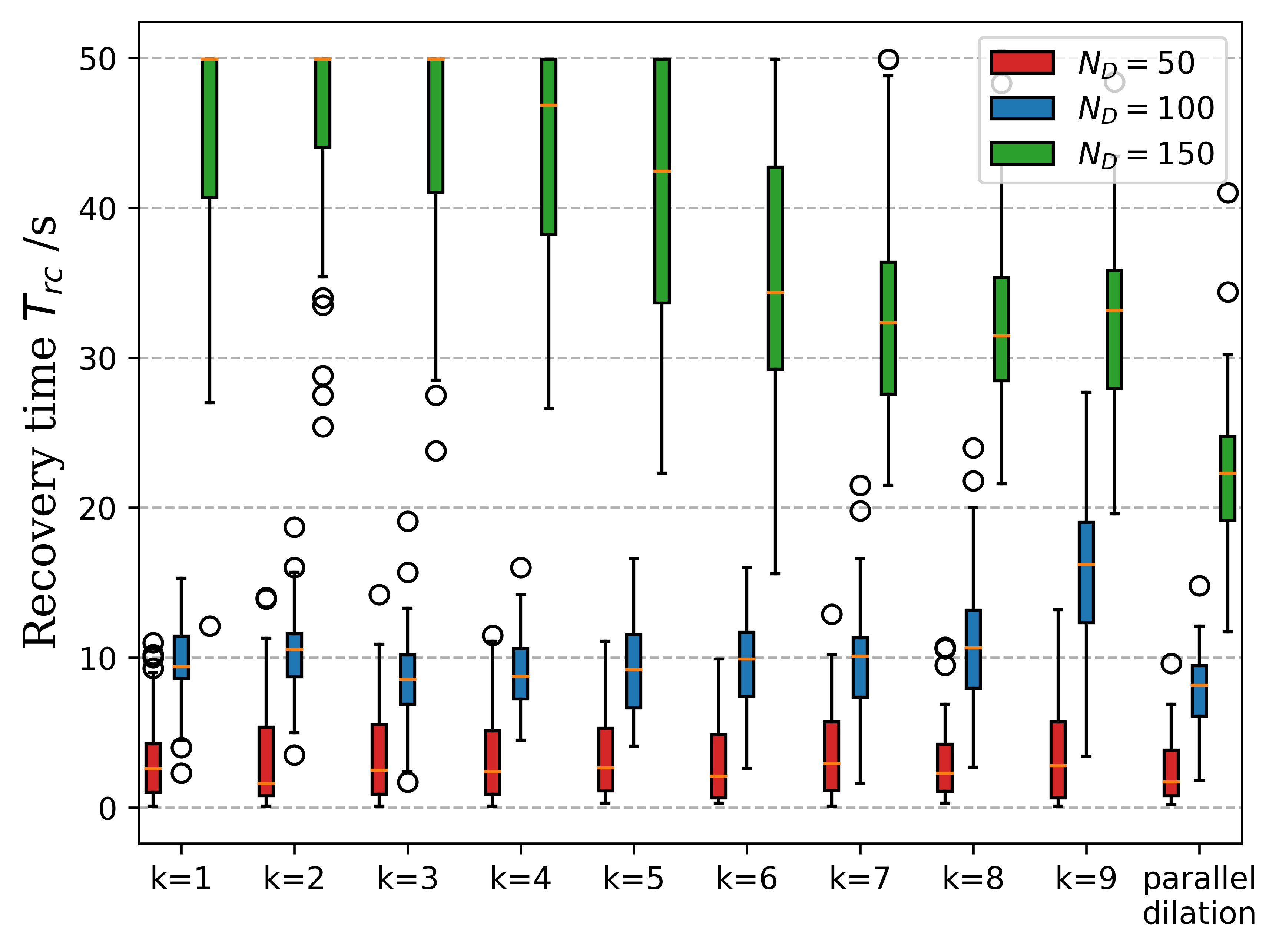}}
\caption{Average recovery time of MBDA with $k$-hop and parallel dilation under different $N_D$.}
\label{batch}
\end{figure}

The proposed MBDA module utilizes $K$ branches for receptive field dilation. 
We illustrate the recovery results of MBDA with a single $k$-th branch or parallel branches in Fig.\ref{batch}. The results indicate that the optimal branches $k$ is dependent on the extent of damage. For example, $k=3$ yields the best performance when $N_D=100$, while $k=9$ is optimal for $N_D=150$. Compared to any fixed dilation size $k$, the parallel dilation structure achieves shorter recovery times across all values of $N_D$. Although this parallel design introduces additional computations, the value of $K$ is limited by (\ref{K}), with acceptable cost considering the substantial performance improvements it provides.

\subsubsection{Study of DGCN}

\begin{figure}[!t]
\centerline{\includegraphics[width=.98\linewidth, height=.74\linewidth]{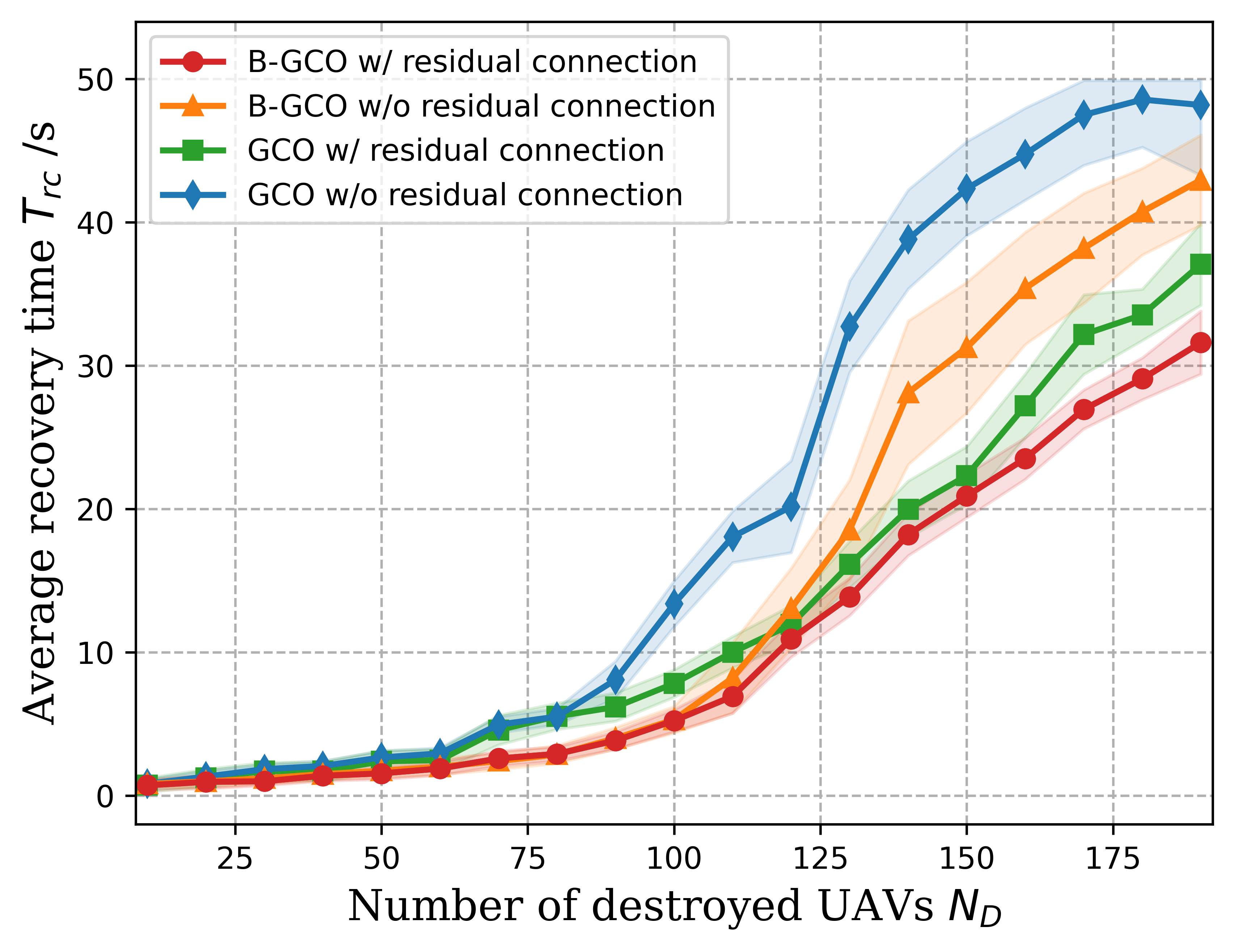}}
\caption{Average recovery time of DGCN with/without key components under different $N_D$.}
\label{ablation}
\end{figure}

The proposed DGCN includes two key components designed to solve CNS problems: the B-GCO and a residual connection mechanism. The average recovery time from ablation experiments on DGCN is shown in Fig.\ref{ablation}, where the B-GCO is replaced by the original GCO from \cite{gcn}, and the residual connection is optionally removed. The results show that omitting the residual connection leads to increased recovery time, indicating reduced convolutional efficiency on sparse topologies. This confirms the effectiveness of the residual connection in handling large-scale damage. In parallel, B-GCO outperforms the original GCO by an average margin of 22.4\%. This improvement stems from their different normalization strategies: the original GCO applies degree normalization at the node level, while B-GCO performs at the graph level to preserve global degree-weight information. Consequently, DGCN can more effectively leverage the original network structure to address CNS challenges.

\subsection{Scalability analysis}

\begin{table*}[!t]
    \centering
    \caption{Statistical results of resilient algorithms under half-damaged scenarios with different swarm scales.}
    \renewcommand\arraystretch{1.8}
    \begin{tabular}{cccc}
    \Xhline{1.5pt}
    Algorithm & \thead{$N=20$ \\ $R_c\,/\,E(T_{rc})\,/\,\sigma(T_{rc})\,/\,E(d)\,/\,max(d)$} & \thead{$N=50$ \\ $R_c\,/\,E(T_{rc})\,/\,\sigma(T_{rc})\,/\,E(d)\,/\,max(d)$} &\thead{$N=100$ \\ $R_c\,/\,E(T_{rc})\,/\,\sigma(T_{rc})\,/\,E(d)\,/\,max(d)$} \\
    \hline
    center-fly & $\textbf{1.00}\,/\,1.95\,/\,1.35\,/\,4.52\,/\,9$ & $\textbf{1.00}\,/\,8.81\,/\,6.13\,/\,10.25\,/\,24$ & $\textbf{1.00}\,/\,11.02\,/\,7.53\,/\,11.73\,/\,49$\\
    HERO & $0.92\,/\,3.93\,/\,4.06\,/\,4.27\,/\,\textbf{8}$ & $0.26\,/\,19.69\,/\,9.32\,/\,4.45\,/\,\textbf{11}$ & $0.08^\dagger\,/\,-\,/\,-\,/-\,/\,-$\\
    SIDR & $0.56\,/\,8.07\,/\,7.11\,/\,\textbf{3.56}\,/\,\textbf{8}$ & $0.70\,/\,9.90\,/\,10.11\,/\,\textbf{3.82}\,/\,\textbf{11}$ & $0.52\,/\,25.93\,/\,12.02\,/\,\textbf{3.97}\,/\,\textbf{11}$\\
    GCN & $\textbf{1.00}\,/\,5.09\,/\,2.81\,/\,5.00\,/\,9$ & $0.98\,/\,12.22\,/\,5.00\,/\,7.38\,/\,20$ & $0.90\,/\,20.52\,/\,9.91\,/\,10.73\,/\,34$\\
    GAT & $\textbf{1.00}\,/\,5.02\,/\,3.63\,/\,4.88\,/\,9$ & $\textbf{1.00}\,/\,10.28\,/\,4.61\,/\,7.74\,/\,20$ & $\textbf{1.00}\,/\,14.83\,/\,5.70\,/\,9.42\,/\,26$\\
    CR-MGC & $\textbf{1.00}\,/\,2.80\,/\,2.59\,/\,4.64\,/\,9$ & $0.98\,/\,9.00\,/\,5.53\,/\,8.44\,/\,23$ & $\textbf{1.00}\,/\,11.14\,/\,5.89\,/\,9.00\,/\,30$\\
    DEMD & $\textbf{1.00}\,/\,1.92\,/\,1.39\,/\,4.49\,/\,9$ & $\textbf{1.00}\,/\,7.78\,/\,5.00\,/\,8.74\,/\,23$ & $\textbf{1.00}\,/\,10.54\,/\,4.91\,/\,8.22\,/\,35$\\
    \hline
    \textbf{ML-DAGL} & $\textbf{1.00}\,/\,\textbf{1.74}\,/\,\textbf{1.19}\,/\,3.97\,/\,9$ & $\textbf{1.00}\,/\,\textbf{3.34}\,/\,\textbf{2.38}\,/\,4.24\,/\,12$ & $\textbf{1.00}\,/\,\textbf{3.72}\,/\,\textbf{1.71}\,/\,4.53\,/\,\textbf{11}$\\
    \Xhline{1.5pt}
    \end{tabular}\\
    \vspace{1em}
    \begin{tabular}{cccccc}
    \Xhline{1.5pt}
    Algorithm & \thead{$N=200$ \\ $R_c\,/\,E(T_{rc})\,/\,\sigma(T_{rc})\,/\,E(d)\,/\,max(d)$} & \thead{$N=500$ \\ $R_c\,/\,E(T_{rc})\,/\,\sigma(T_{rc})\,/\,E(d)\,/\,max(d)$} &\thead{$N=1000$ \\ $R_c\,/\,E(T_{rc})\,/\,\sigma(T_{rc})\,/\,E(d)\,/\,max(d)$} \\
    \hline
    center-fly & $0.98\,/\,22.90\,/\,11.17\,/\,26.54\,/\,98$ & $0.88\,/\,46.70\,/\,18.81\,/\,71.04\,/\,238$ & $0.50\,/\,90.12\,/\,26.16\,/\,260.78\,/\,463$\\
    HERO & $0^\dagger\,/\,-\,/\,-\,/\,-\,/\,-$ & $0^\dagger\,/\,-\,/\,-\,/\,-\,/\,-$ & $0^\dagger\,/\,-\,/\,-\,/\,-\,/\,-$\\
    SIDR & $0.36\,/\,34.81\,/\,20.43\,/\,\textbf{4.53}\,/\,16$ & $0.32\,/\,69.94\,/\,14.81\,/\,\textbf{4.19}\,/\,\textbf{13}$ & $0.22\,/\,90.64\,/\,41.40\,/\,\textbf{4.39}\,/\,\textbf{15}$\\
    GCN & $0.86\,/\,30.12\,/\,12.72\,/\,12.14\,/\,63$ & $0.86\,/\,58.52\,/\,14.52\,/\,56.72\,/\,214$ & $0.08^\dagger\,/\,-\,/\,-\,/\,-\,/\,-$\\
    GAT & $\textbf{1.00}\,/\,24.48\,/\,9.64\,/\,13.32\,/\,65$ & $0.94\,/\,56.90\,/\,12.98\,/\,21.81\,/\,96$ & $0.80\,/\,95.83\,/\,13.50\,/\,38.12\,/\,219$\\
    CR-MGC & $\textbf{1.00}\,/\,21.28\,/\,8.68\,/\,17.66\,/\,84$ & $0.76\,/\,63.84\,/\,14.34\,/\,29.39\,/\,108$ & $0.34\,/\,104.13\,/\,15.62\,/\,33.38\,/\,188$\\
    DEMD & $\textbf{1.00}\,/\,21.40\,/\,7.32\,/\,13.37\,/\,53$ & $\textbf{1.00}\,/\,45.40\,/\,8.72\,/\,18.14\,/\,67$ & $0.68\,/\,86.89\,/\,22.79\,/\,25.77\,/\,140$\\
    \hline
    \textbf{ML-DAGL} & $\textbf{1.00}\,/\,\textbf{5.24}\,/\,\textbf{2.10}\,/\,5.24\,/\,\textbf{15}$ & $\textbf{1.00}\,/\,\textbf{18.10}\,/\,\textbf{7.69}\,/\,7.46\,/\,30$ & $\textbf{1.00}\,/\,\textbf{32.58}\,/\,\textbf{6.18}\,/\,8.53\,/\,39$\\
    \Xhline{1.5pt}
    \end{tabular}\\
    \vspace{0.3em}
    \leftline{$^\dagger$Results with $R_c < 0.1$ indicate that the algorithm fails to converge in most cases so that other metrics are not statistically significant.}
    \label{scalability}
\end{table*}

To evaluate the scalability of the proposed algorithm for varying swarm scales, we set the number of UAVs as $N=50,100,200,500,1000$. The damage ratio is set to $p=0.5$, and each simulation of different $N$ is repeated 50 times randomly for the statistical results. Simulation results for resilience algorithms are summarized in TABLE~\ref{scalability}, including the convergent ratio $R_c$, mean recovery time $E(T_{rc})$, variance of recovery time $\sigma(T_{rc})$, average node degree $E(d)$, and maximum node degree $\max(d)$. 

The results show that ML-DAGL consistently maintains $R_c=1$ with full connectivity restoration across different swarm sizes, demonstrating strong scalability. In contrast, other algorithms exhibit declining $R_c$ as the swarm size increases, especially with poor performance when $N$ reaches 1000. Meanwhile, regarding recovery efficiency, ML-DAGL achieves state-of-the-art results in both the mean and variance of recovery time, enabling more efficient and stable recovery solutions. For metrics on degree distribution, the ML-DAGL yields the second lowest values for both $E(d)$ and $\max(d)$, only inferior to the distributed algorithm SIDR. In summary, the proposed algorithm has excellent scalability.

\section{Conclusion}
In this paper, we studied the CNS issue of the USNET under massive damage scenarios. To cope with the challenges of utilizing graph-learning approaches to generate optimal recovery trajectories, we proposed a novel ML-DAGL algorithm. 
To adapt to diverse damage cases with varying scales, we introduced the MBDA module to form a sequence of mDAGs with different ranges of receptive fields. 
Each mDAG maintains the edges between remaining and destroyed nodes within the k-hop range, establishing dilated topologies with damage attention.
We then developed the DGCN to generate the optimal recovery trajectories based on the mDAG sequence, utilizing the bipartite GCO and residual connection to enhance the feature mining and aggregation. Additionally, we theoretically proved the convergence of the proposed algorithm and analyzed the upper bound of recovery time and computational complexity. The proposed algorithm is feasible for practical deployment with limited memory requirement. Simulations demonstrated that ML-DAGL algorithm have excellent scalability, along with significant improvements in convergence, recovery time, and topology distribution. 


{\appendices
\section{Proof of Proposition \ref{prop-c2}}
In the metric space of position matrices $\{\bm{X}_{gco}\}\subseteq\mathbb{R}^{N\times 2}$, we can represent the matrix form of B-GCO $\bm{X}_{gco}^{l+1}=g_\theta\ast \bm{X}_{gco}^l=(\bm{I}_N-\epsilon\bm{L}_{dag}^k)\bm{X}_{gco}^l$ as
\begin{align}
    \begin{pmatrix}
    x_{1}^{l+1} & y_{1}^{l+1} \\
    x_{2}^{l+1} & y_{2}^{l+1} \\
    \vdots & \vdots \\
    x_{N}^{l+1} & y_{N}^{l+1}
    \end{pmatrix}=\begin{pmatrix}
    l_{11}^k &\!\!\!\! l_{12}^k &\!\!\!\! \ldots &\!\!\!\! l_{1N}^k \\
    l_{21}^k &\!\!\!\! l_{22}^k &\!\!\!\! \ldots &\!\!\!\! l_{2N}^k \\
    \vdots &\!\!\!\! \vdots &\!\!\!\! \ddots &\!\!\!\! \vdots \\
    l_{N1}^k &\!\!\!\! l_{N2}^k &\!\!\!\! \ldots &\!\!\!\! l_{NN}^k
    \end{pmatrix}\begin{pmatrix}
    x_{1}^l & y_{1}^l \\
    x_{2}^l & y_{2}^l \\
    \vdots & \vdots \\
    x_{N}^l & y_{N}^l
    \end{pmatrix},
\end{align}
where $x_{i}$ and $y_{i}$ denote the coordinate component of $X$ and $Y$ axis from the position vector $\bm{p}_{i}$ of UAV $u_{i}$, and $l_{ij}^k$ is the element of $\bm{I}_N-\epsilon\bm{L}_{dag}^k$. Denote $d_{i}^k=\sum_{j=1}^Na_{ij}^k$ as the degree of node $u_{i}$ from the mDAG, we have
\begin{align}
\mathop{\sum}\limits_{i=1}^Nl_{ij}^k=1+\epsilon d_{i}^k-\epsilon\mathop{\sum}\limits_{j=1}^Na_{ij}^k=1.
\end{align}

The B-GCO can guarantee the invariance of column sums, i.e.,
\begin{align}
\mathop{\sum}\limits_{i=1}^Nx_{i}^{l+1}=\mathop{\sum}\limits_{i=1}^N\mathop{\sum}\limits_{j=1}^Nl_{ij}^kx_{j}^l=\mathop{\sum}\limits_{j=1}^N(x_{j}^l\mathop{\sum}\limits_{i=1}^Nl_{ij}^k)=\mathop{\sum}\limits_{j=1}^Nx_{j}^{l}\ ,
\end{align}
and the equality $\sum_{i=1}^Ny_{i}^{l+1}=\sum_{j=1}^Ny_{j}^l$ can be proved in the same manner.

Therefore, the B-GCO is closed for the metric space $\{\bm{X}_{gco}|\sum_{i=1}^Nx_{i}=X_{sum},\sum_{i=1}^Ny_{i}=Y_{sum}\}$, which means every position matrix $\bm{X}_{gco}^l$ share a same central location $\bm{p}_c=\frac{1}{N}\sum_{i=1}^N\bm{p}_{i}=[\frac{1}{N}X_{sum},\frac{1}{N}Y_{sum}]^\top$ from $\bm{X}_{gco}$.

Then we use the \textit{Contraction Mapping Principle} to prove the B-GCO is a contraction operation and the position matrix $\bm{X}_{gco}$ is convergent to a certain matrix $\bm{\bar{X}}_{gco}$. We first define the distance between two position matrices $\bm{X}_{gco}^a$ and $\bm{X}_{gco}^b$ as 
\begin{equation}
\begin{aligned}
    d(\bm{X}_{gco}^a,\bm{X}_{gco}^b)&=\|\bm{X}_{gco}^a-\bm{X}_{gco}^b\|_\infty\\
    &=\mathop{\rm max}\limits_{1\leq i\leq N}\{|x_{i}^a-x_{i}^b|+|y_{i}^a-y_{i}^b|\}.
\end{aligned}
\end{equation}

The distance between the B-GCO of $\bm{X}_{gco}^a$ and $\bm{X}_{gco}^b$ is then calculated as
\begin{equation}
\begin{aligned}
    d(g_\theta\!\ast\!\bm{X}_{gco}^a,g_\theta\!\ast\!\bm{X}_{gco}^b)&=\|g_\theta\!\ast\!\bm{X}_{gco}^a-g_\theta\!\ast\!\bm{X}_{gco}^b\|_\infty\\
    &=\|(\bm{I}_N\!-\!\epsilon\bm{L}_{dag}^k)(\bm{X}_{gco}^a\!-\!\bm{X}_{gco}^b)\|_\infty.
\end{aligned}
\end{equation}

Since the infinity norm of matrix $\|\cdot\|\infty$ has the sub-multiplicative property, we can get
\begin{equation}
\begin{aligned}
    d(g_\theta\!\ast\!\bm{X}_{gco}^a,g_\theta\!\ast\!\bm{X}_{gco}^b)\leq \|\bm{I}_N\!-\!\epsilon\bm{L}_{dag}^k\|_\infty\|\bm{X}_{gco}^a\!-\!\bm{X}_{gco}^b\|_\infty&\\
    =\mathop{\rm max}\limits_{1\leq i\leq N}\{|1\!-\!\epsilon d_{i}^k|\!+\!\mathop{\sum}\limits_{j=1}^N\epsilon a_{ij}^k\}\|\bm{X}_{gco}^a\!-\!\bm{X}_{gco}^b\|_\infty&.
\end{aligned}
\end{equation}

Notice that $d_{ij}^k\leq\|\bm{A}_{dag}^k\|_\infty$, when $0<\epsilon\leq\frac{1}{\|\bm{A}_{dag}^k\|_\infty}$, there is
\begin{align}
    1-\epsilon d_{i}^k\geq1-\epsilon\|\bm{A}_{dag}^k\|_\infty\geq 1-\frac{1}{\|\bm{A}_{dag}^k\|_\infty}\|\bm{A}_{dag}^k\|_\infty=0,
\end{align}
hence we have
\begin{equation}
\begin{aligned}
    &d(g_\theta\ast \bm{X}_{gco}^a,g_\theta\ast \bm{X}_{gco}^b)\\
    &\leq\mathop{\rm max}\limits_{1\leq i\leq N}\{|1-\epsilon d_{i}^k|+\mathop{\sum}\limits_{j=1}^N\epsilon a_{ij}^k\}\|\bm{X}_{gco}^a-\bm{X}_{gco}^b\|_\infty\\
    &=\mathop{\rm max}\limits_{1\leq i\leq N}\{1-\epsilon d_{i}^k+\mathop{\sum}\limits_{j=1}^N\epsilon a_{ij}^k\}\|\bm{X}_{gco}^a-\bm{X}_{gco}^b\|_\infty\\
    &=\mathop{\rm max}\limits_{1\leq i\leq N}\{1-\epsilon \mathop{\sum}\limits_{j=1}^Na_{ij}^k+\mathop{\sum}\limits_{j=1}^N\epsilon a_{ij}^k\}\|\bm{X}_{gco}^a-\bm{X}_{gco}^b\|_\infty\\
    &=\mathop{\rm max}\limits_{1\leq i\leq N}\{1\}\|\bm{X}_{gco}^a-\bm{X}_{gco}^b\|_\infty\\
    &=d(\bm{X}_{gco}^a,\bm{X}_{gco}^b).
    \label{dis}
\end{aligned}
\end{equation}

The condition for (\ref{dis}) to take the equal sign is
\begin{equation}
\begin{aligned}
    &\|(\bm{I}_N\!-\!\epsilon\bm{L}_{dag}^k)(\bm{X}_{gco}^a\!-\!\bm{X}_{gco}^b)\|_\infty\\=&
    \|\bm{I}_N\!-\!\epsilon\bm{L}_{dag}^k\|_\infty\|\bm{X}_{gco}^a\!-\!\bm{X}_{gco}^b\|_\infty.
\end{aligned}
\end{equation}
Given the proof of the sub-multiplicative property of $\|\cdot\|\infty$ as
\begin{equation}
\begin{aligned}
    &\|(\bm{I}_N-\epsilon\bm{L}_{dag}^k)(\bm{X}_{dag}^a-\bm{X}_{dag}^b)\|_\infty\\
    &=\mathop{\rm max}\limits_{1\leq i\leq N}\{|\mathop{\sum}\limits_{j=1}^Nl_{ij}^k(x_{j}^a-x_{j}^b)|+|\mathop{\sum}\limits_{j=1}^Nl_{ij}^k(y_{j}^a-y_{j}^b)|\}\\
    &\leq\mathop{\rm max}\limits_{1\leq i\leq N}\{\mathop{\sum}\limits_{j=1}^N|l_{ij}^k(x_{j}^a-x_{j}^b)|+\mathop{\sum}\limits_{j=1}^N|l_{ij}^k(y_{j}^a-y_{j}^b)|\}\\
    &=\mathop{\rm max}\limits_{1\leq i\leq N}\{\mathop{\sum}\limits_{j=1}^N|l_{ij}^k|\cdot(|x_{j}^a-x_{j}^b|+|y_{j}^a-y_{j}^b|)\}\\
    &\leq \mathop{\rm max}\limits_{1\leq i\leq N}\mathop{\sum}\limits_{j=1}^N|l_{ij}^k|\cdot(\mathop{\rm max}\limits_{1\leq j\leq N}\{|x_{j}^a-x_{j}^b|+|y_{j}^a-y_{j}^b|\})\\
    &=\|\bm{I}_N-\epsilon\bm{L}_{dag}^k\|_\infty\|\bm{X}_{gco}^a-\bm{X}_{gco}^b\|_\infty.
    \label{pf}
\end{aligned}
\end{equation}
we can see that inequality (\ref{pf}) is scaled for twice. The first scaling is based on the triangle inequality, hence the condition to take the equal sign is $l_{ij}^k(x_{j}^a-x_{j}^b)\geq0$ and $l_{ij}^k(y_{j}^a-y_{j}^b)\geq0$ for every $1\leq j\leq N$. Since $l_{ij}^k$ is always non-negative when $0<\epsilon\leq\frac{1}{\|\bm{A}_{dag}^k\|_\infty}$, i.e.,
\begin{equation}
\begin{aligned}
l_{ij}^k=
\left\{ 
    \begin{array}{lc}
    1-\epsilon d_{i}^k\geq 0,&i=j;\\
    \epsilon a_{ij}^k\geq 0,&i\neq j,
    \end{array}
\right.
\end{aligned}
\end{equation}
we can get the first scaling to be an equality requires
\begin{align}
\left\{ 
    \begin{array}{lc}
    x_{j}^a-x_{j}^b\geq0\\
    y_{j}^a-y_{j}^b\geq0
    \end{array}
\right.1\leq j\leq N.
\label{c1}
\end{align}

The condition for the second scaling to be equal is 
\begin{align}
    |x_{j}^a-x_{j}^b|+|y_{j}^a-y_{j}^b|=\|\bm{X}_{gco}^a-\bm{X}_{gco}^b\|_\infty=C\  \text{for}\  \forall j,
\end{align}
where $C\in\mathbb{R}$ is a constant.

Assuming that the two position matrices $\bm{X}_{gco}^a,\bm{X}_{gco}^b\in\{\bm{X}_{gco}|\sum_{i=1}^Nx_{i}=X_{sum},\sum_{i=1}^Ny_{i}=Y_{sum}\}$, we can derive that
\begin{equation}
\begin{aligned}
    C&=\frac{1}{N}\mathop{\sum}\limits_{j=1}^N|x_{j}^a-x_{j}^b|+|y_{dj}^a-y_{j}^b| \\
    &=\frac{1}{N}\mathop{\sum}\limits_{j=1}^N(x_{j}^a-x_{j}^b+y_{j}^a-y_{j}^b) \\
    &=\frac{1}{N}(\mathop{\sum}\limits_{j=1}^Nx_{j}^a-\mathop{\sum}\limits_{j=1}^Nx_{j}^b+\mathop{\sum}\limits_{j=1}^Ny_{j}^a-\mathop{\sum}\limits_{j=1}^Ny_{j}^b) \\
    &=\frac{1}{N}(X_{sum}-X_{sum}+Y_{sum}-Y_{sum})\\
    &=0.
\end{aligned}
\end{equation}
This indicates that $x_{j}^a-x_{j}^b=0$ and $y_{j}^a-y_{j}^b=0$ for every $1\leq j\leq N$. Hence, (\ref{dis}) is an equality if and only if $\bm{X}_{gco}^a=\bm{X}_{gco}^b$, and we have
\begin{align}
    d(g_\theta\ast \bm{X}_{gco}^a,g_\theta\ast \bm{X}_{gco}^b)=d(\bm{X}_{gco}^a,\bm{X}_{gco}^b)=0.
\end{align}

At this point, we have proved that for $\forall\bm{X}_{gco}^a,\bm{X}_{gco}^b\in\{\bm{X}_{gco}|\sum_{i=1}^Nx_{i}=X_{sum},\sum_{i=1}^Ny_{i}=Y_{sum}\}$, the condition
\begin{align}
    d(g_\theta\ast \bm{X}_{gco}^a,g_\theta\ast \bm{X}_{gco}^b)\leq\alpha d(\bm{X}_{gco}^a,\bm{X}_{gco}^b)
\end{align}
always holds for $\alpha\in(0,1)$. This indicates that the B-GCO is a contraction operation when $0<\epsilon\leq\frac{1}{\|\bm{A}_{dag}^k\|_\infty}$, and there exists and only exists one position matrix $\bm{\bar{X}}_{gco}=[\bm{\bar{p}}_{1},\bm{\bar{p}}_{r_{2}},...,\bm{\bar{p}}_{N}]^\top$ such that
\begin{align}
    \bm{\bar{X}}_{gco}=(\bm{I}_N-\epsilon\bm{L}_{dag}^k)\bm{\bar{X}}_{gco}=\mathop{\rm lim}\limits_{l\rightarrow\infty}(\bm{I}_N-\epsilon\bm{L}_{dag}^k)^l\bm{X}_{gco},
\end{align}
which is also called the \textit{Banach point}. 

By eliminating the $\bm{\bar{X}}_d$ on both side, we can derive
\begin{align}
    \bm{L}_{dag}^k\bm{\bar{X}}_{gco}=\bm{L}_{dag}^k[\bm{\bar{x}},\bm{\bar{y}}]=\bm{0},
\end{align}
where $\bm{\bar{x}}$ and $\bm{\bar{y}}$ are the column vectors in $\bm{\bar{X}}_{gco}$. This indicates that $\bm{\bar{x}}$ and $\bm{\bar{y}}$ are eigenvectors of $\bm{L}_{dag}^k$ corresponding to the zero eigenvalue. Note that the number of zero eigenvalues of $\bm{L}_{dag}^k$ equals the number of sub-nets in the mDAG, and $S$ zero eigenvalues $\lambda_1=0,\lambda_2=0,...,\lambda_S=0$ correspond to $S$ orthogonal eigenvectors $\{\bm{u}_1,\bm{u}_2,...,\bm{u}_S\}$. Thereby, we can represent $\bm{\bar{x}}$ and $\bm{\bar{y}}$ as the linear combinations of $\bm{u}_i$, i.e.,
\begin{equation}
\begin{aligned}
    \left\{ 
    \begin{array}{lc}
    \bm{\bar{x}}_d=\mathop{\sum}\limits_{i=1}^S\alpha_i\bm{u}_i;\\
    \bm{\bar{y}}_d=\mathop{\sum}\limits_{i=1}^S\beta_i\bm{u}_i,
    \end{array}
\right.
\label{cm}
\end{aligned}
\end{equation}
where $\alpha_i,\beta_i\in\mathbb{R}$ are coefficients of the combinations.

When the mDAG is connected, the number of zero eigenvalues of $\bm{L}_{dag}^k$ equals 1. Note that the vector $\bm{1}_N$ must be a eigenvector of $\bm{L}_{dag}^k$ corresponding to the zero eigenvalue, since
\begin{equation}
\begin{aligned}
    \bm{L}_{dag}^k\bm{1}_N=\begin{pmatrix}
    d_{1}-\sum_{j=1}^Na_{1j}\\
    d_{2}-\sum_{j=1}^Na_{2j}\\
    \vdots\\
    d_{N}-\sum_{j=1}^Na_{Nj}
    \end{pmatrix}=\bm{0}=0\cdot\bm{1}_N.
\end{aligned}
\end{equation}

Based on (\ref{cm}), we can get $\bm{\bar{x}}=\alpha_1\bm{1}_N$ and $\bm{\bar{y}}=\beta_1\bm{1}_N$ with $\alpha_1,\beta_1\neq0$. Since $\bm{\bar{X}}_{gco}\in\{\bm{X}_{gco}|\sum_{i=1}^Nx_{i}=X_{sum},\sum_{i=1}^Ny_{i}=Y_{sum}\}$, we have
\begin{align}
    \mathop{\sum}\limits_{i=1}^Nx_{i}=\mathop{\sum}\limits_{i=1}^N\alpha_1
    \Rightarrow\alpha_1=\frac{1}{N}X_{sum},\\
    \mathop{\sum}\limits_{i=1}^Ny_{i}=\mathop{\sum}\limits_{i=1}^N\beta_1
    \Rightarrow\beta_1=\frac{1}{N}Y_{sum}.
\end{align}
Therefore, when the input dilated graph is connected with the feature matrix as $\bm{X}_{in}$, $\bm{\bar{X}}_{gco}\equiv[\bm{p}_c,\bm{p}_c,...,\bm{p}_c]^\top$ holds. This completes the proof.

\section{Proof of Proposition \ref{prop-cp}}
The DGCN consists of an input normalization layer, $Q$ B-GCO layers, $Q-1$ nonlinear activation layers, and an output upscale layer. Compared to other operations, the graph convolution within the B-GCO layers takes the most computational complexity. For each B-GCO layer, the GCO applies matrix multiplications for two times in (\ref{w}). The computational complexity of a single B-GCO layer is the sum of the complexities of matrix multiplications.

For the $q$-th B-GCO layer, the first multiplication is the GCO kernel $g_\theta=\bm{I}_{KN}-\epsilon\bm{\dot{L}}_{dag}$, shaped ${KN\times KN}$, multiplying with the feature matrix $\bm{\dot{X}}_{gco}^{q-1}$ with shape of ${KN\times d_{in}^q}$, where $d_{in}^q$ is the input dimension of the $q$-th layer. Notice that the GCO kernel is a block-diagonal matrix calculated by $\bm{\dot{A}}_{dag}$, and each $\bm{A}_{dag}^k$ has at least $\frac{1}{2}N^2$ zero elements according to (\ref{Hadamard}). Therefore, the GCO kernel is a sparse matrix, as its ratio of the total number of non-zero elements to the total number of all elements in the matrix is
\begin{align}
    \eta \leq \frac{K\cdot\frac{1}{2}N^2}{KN\cdot KN}=\frac{1}{2K}.
\end{align}
Therefore, the computational complexity of the first multiplication is $\mathcal{O}\left(\text{nnz}(\bm{I}_{KN}\!-\!\epsilon\bm{\dot{L}}_{dag})\cdot d_{in}^q\right)$. 

The number of non-zero elements in the B-GCO kernel is the sum of the numbers of non-zero elements in each branch with $\bm{A}_{dag}^k$. Let's first analyze the 1-hop branch. The element $a_{r_id_j}^1=1$ in $\bm{A}_{dag}^1$ indicates that the remaining node $u_{r_i}$ and the destroyed nodes $u_{d_j}$ established a communication in the original USNET. Assuming that $N$ nodes are randomly distributed within a $D\times D$ area, the node degree follows a Poisson distribution with an average degree $\bar{d}$. Denote $p=N_D/N$ as the damage ratio, the number of communication links between the remaining nodes and destroyed nodes is represented as
\begin{equation}
\begin{aligned}
    \text{nnz}(\bm{A}_{in,r_id_j}^1)=&N_R\cdot \bar{d}_R\\=&(1-p)N\cdot p\bar{d}\\=&p(1-p)\pi d_{tr}^2\rho N,
\end{aligned}
\end{equation}
where $d_{tr}$ is the maximum communication range determined by (\ref{dtr}), $\rho=N/D^2$ is the node density. Therefore, the number of non-zero elements in $\bm{A}_{dag}^1$ is $\text{nnz}(\bm{A}_{dag}^k)=2p(1-p)\pi d_{tr}^2\rho N$ based on (\ref{Hadamard}). For the $k$-hop branch, the node degree still follows the Poisson distribution with the mean as $k\bar{d}$, hence we have $\text{nnz}(\bm{A}_{dag}^k)=k\cdot\text{nnz}(\bm{A}_{dag}^1)$. Since $\text{nnz}(\bm{I}_N\!-\!\epsilon\bm{L}_{dag}^k)=\text{nnz}(\bm{A}_{dag}^k)\!+\!N$, we have the number of non-zero elements in the B-GCO kernel as
\begin{equation}
\begin{aligned}
    \text{nnz}(\bm{I}_N\!-\!\epsilon\bm{\dot{L}}_{dag})=&\mathop{\sum}\limits_{k=1}^K\text{nnz}(\bm{A}_{dag}^k)+KN\\=&\mathop{\sum}\limits_{k=1}^K k\cdot 2p(1-p)\pi d_{tr}^2\rho N+KN\\
    =&p(1-p)\pi d_{tr}^2\rho K(K-1)N+KN.
\end{aligned}
\end{equation}
The second multiplication in the $q$-th layer is the middle matrix with shape $KN\times d_{in}^q$ multiplying with the weight matrix $\bm{W}^q$, shaped $d_{in}^q\times d_{out}^q$. The computational complexity of the second multiplication is $\mathcal{O}\left(KNd_{in}^qd_{out}^q\right)$. Therefore, the complexity of the $q$-th layer is $\mathcal{O}\left(\text{nnz}(\bm{I}_{KN}\!-\!\epsilon\bm{\dot{L}}_{dag})\!\cdot\! d_{in}^q\!+\!KNd_{in}^qd_{out}^q\right)$.

The total complexity of DGCN is the sum of the B-GCO layers, i.e.,
\begin{equation}
\begin{aligned}
&\mathcal{O}\left(\mathop{\sum}\limits_{q=1}^Q\text{nnz}(\bm{I}_{KN}\!-\!\epsilon\bm{\dot{L}}_{dag})\cdot d_{in}^q\!+\!KNd_{in}^qd_{out}^q\right)\\
=&\mathcal{O}\!\left(\mathop{\sum}\limits_{q=1}^Q[p(1\!-\!p)\pi d_{tr}^2\rho K(K\!-\!1)N\!+\!K\!N]d_{in}^q\!+\!K\!Nd_{in}^qd_{out}^q\!\right)\\
=&\mathcal{O}\left(KN\mathop{\sum}\limits_{q=1}^Q[p(1\!-\!p)\pi d_{tr}^2\rho(K\!-\!1)\!+\!1\!+\!d_{out}^q]d_{in}^q\right).
\end{aligned}
\end{equation}
This completes the proof.
}

 


\bibliographystyle{IEEEtran}
\balance
\bibliography{IEEEabrv,main}


 




\vfill

\end{document}